\documentclass[a4paper,fleqn,useAMS,usenatbib,english]{mnras}
\usepackage{amsmath,amssymb,amstext}
\usepackage{newtxtext,newtxmath}
\usepackage{graphicx}
\usepackage[T1]{fontenc}
\usepackage{ae,aecompl}
\usepackage[figure,figure*]{hypcap}
\usepackage{xspace}
\usepackage{xcolor}
\usepackage[nameinlink,capitalize]{cleveref}
\usepackage{comment}

\def\equationautorefname~#1\null{equation~(#1)\null}

\global\long\def\hMpc{h^{-1}\mathrm{Mpc}}
\global\long\def\bx{{\mathbf{x}}}
\global\long\def\d{\mathrm{d}}
\global\long\def\av#1{\left\langle #1\right\rangle }
\global\long\def\vr{\varrho_{0}}

\global\long\def\hDens{h^{3}\mathrm{Mpc}^{-3}}
\crefname{figure}{Figure}{Figures}
\crefname{equation}{equation}{equations}

\begin{document}
\label{firstpage}
\pagerange{\pageref{firstpage}--\pageref{lastpage}}

\title[Redshift evolution of topological clustering]{The clustering of galaxies in the SDSS-III Baryon Oscillation Spectroscopic Survey: Evolution of higher-order correlations demonstrated with Minkowski Functionals}
	
\author[J. M. Sullivan et al.]{
James M. Sullivan,$^{1,2}$\thanks{E-mail: jsull3@utexas.edu}
Alexander Wiegand,$^{1,3}$
Daniel J. Eisenstein$^{1}$
\\
$^{1}$Harvard-Smithsonian Center for Astrophysics, 60 Garden St, Cambridge, MA, 02138, USA\\
$^{2}$University of Texas at Austin, 110 Inner Campus Dr, Austin, TX 78705, USA\\
$^{3}$Max--Planck--Institut f\"ur Astrophysik, Karl--Schwarzschild--Str.~1, D--85741 Garching, Germany
}
\pubyear{2017}

\maketitle

\begin{abstract}
We probe the higher-order galaxy clustering in the final data release (DR12) of the Sloan Digital Sky Survey Baryon Oscillation Spectroscopic Survey (BOSS) using germ-grain Minkowski Functionals (MFs). Our data selection contains $979,430$ BOSS galaxies from both the northern and southern galactic caps over the redshift range $z=0.2 - 0.6$. We extract the higher-order part of the MFs, detecting the deviation from the purely Gaussian case with $\chi^2 \sim \mathcal{O}(10^3)$ on 24 degrees of freedom across the entire data selection.
We measure significant redshift evolution in the higher-order functionals for the first time. We find $15-35\%$ growth, depending on functional and scale, between our redshift bins centered at $z=0.325$ and $z=0.525$. We show that the structure in higher order correlations grow faster than that in the two-point correlations, especially on small scales where the excess approaches a factor of $2$. We demonstrate how this trend is generalizable by finding good agreement of the data with a hierarchical model in which the higher orders grow faster than the lower order correlations. We find that the non-Gaussianity of the underlying dark matter field grows even faster than the one of the galaxies due to decreasing clustering bias. Our method can be adapted to study the redshift evolution of the three-point and higher functions individually.
\end{abstract}


\section{Introduction}\label{sec:intro}

Cosmic large-scale structure is largely understood by characterizing and modeling observed galaxy distributions, which strongly influence the accepted cosmological model. Recent galaxy redshift surveys contain spectroscopic redshifts for of order $10^{5} - 10^{6}$ galaxies, and their size makes them ideal testing grounds for exploring non-Gaussian features in this structure. Any quantitative analysis of such features must be firmly rooted in precise statistical measures. The standard analysis employs the two-point correlation function $\xi_{2}$, but the condensed information offered by this function fails to fully account for fundamental observed structures such as sheets and filaments.
To accurately describe large-scale structure we must then look beyond two-point statistics to higher-order correlation information. 

The simplest approach is to calculate the higher-order correlation functions directly. While this has been widely achieved for the two-point function, and recently quite successfully for the three-point function (e.g. by \citet{2015MNRAS.454.4142S} and \citet{2017MNRAS.469.1738S}), fourth- and higher-order functions have yet to be well-determined. As it turns out, computing these functions directly is computationally infeasible. 
There are various alternatives, and one of the most useful and rigorous is Minkowski Functional (MF) analysis. MFs quantitatively describe the geometry of extended bodies by mapping the shape of a body to the real numbers. These functionals uniquely characterize the geometry and topology of a galaxy distribution, and contain information about all higher-order correlation functions. 

MFs were first used to characterize large-scale structure by \citet{1994A&A...288..697M} in the form of the germ-grain model. This model pins down the morphology of the galaxy distribution by treating the survey galaxies as points (the germs) and decorating them with balls (the grains) whose scale-probing radius is the only model parameter. The union of these balls creates a set of extended bodies to which methods from integral geometry can be applied (For a review see \citet{1999elss.conf..127S} or \citet{1996dmu..conf..281S}, or \citet{1995lssu.conf..156B} for a short review.). We discuss this model further in the context of our analysis. 

Another popular use of MFs is to apply them to the isodensity contours of density fields, including galaxy and cluster surveys (e.g. \citet{1996dmu..conf..281S}, \citet{1997MNRAS.284...73K}, \citet{1998A&A...333....1K}, \cite{2001A&A...373....1K}, and \citet{2001A&A...377....1K}), dark matter overdensity fields (\citet{1996app..conf..251P}, \citet{1997ApJ...482L...1S}, \citet{1998ApJ...495L...5S}, \mbox{\citet{1998ApJ...508..551S}}, \mbox{\citet{1999ApJ...526..568S}}, \citet{2003PASJ...55..911H}, \citet{2004astro.ph..8428N}, \citet{2013ApJS..209...19C}, and \citet{2014MNRAS.437.2488B}), and other astrophysical settings (\citet{2013PhRvD..88l3002P}, \citet{2006MNRAS.370.1329G}, \citet{2014A&A...562A..87E}, and \citet{2017MNRAS.465..394Y}). Still further work with MFs has recently used the isotemperature contour maps of the CMB to constrain its Gaussianity (\citet{2013MNRAS.429.2104D}, \citet{2014A&A...571A..23P}, and \citet{2014A&A...571A..25P}, \citet{2016A&A...594A..16P}). Clearly MFs are a 
widely-used tool in a variety of cosmological subfields. 

\begin{figure}
\centering
\includegraphics[width=0.45\textwidth]{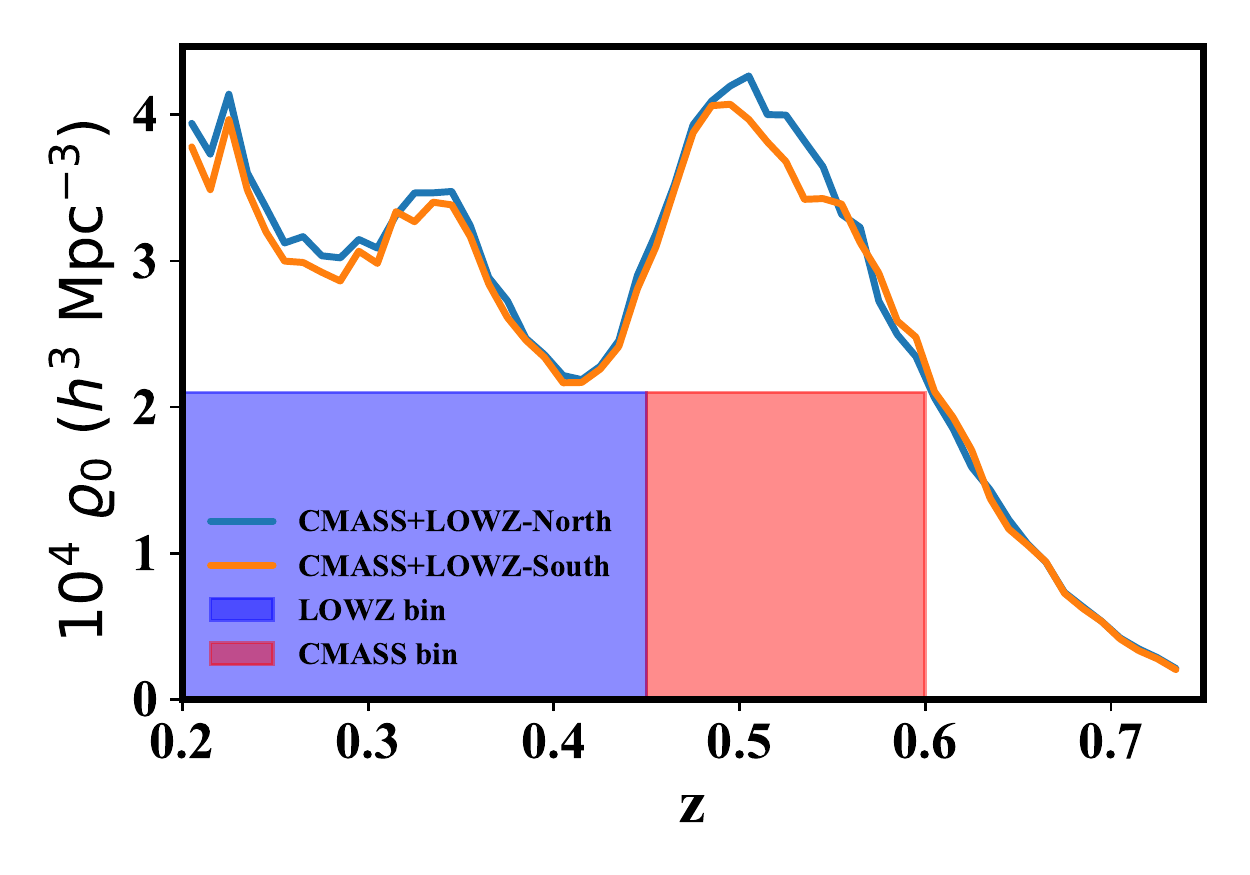}
\caption{Minimum number density for the combined full CMASS, LOWZ, and LOWZE2 catalogs. By considering only values of $\vr$ that are at or below the minimum value over the combined sample, we need not restrict our analysis to differing density ranges at different redshifts. The bins we call CMASS and LOWZ in this work are shown here.}
\label{fig:bins}
\end{figure}

Here we focus on the germ-grain model, because recent work using the 7th and 12th data releases of SDSS-III (DR7 and DR12) of the Baryon Oscillation Spectroscopic Survey (BOSS - \citet{2013AJ....145...10D}) has shown its use in accessing correlation information that cannot be calculated directly {(DR7 paper - \citet{2014MNRAS.443..241W}, \hypertarget{txt:dr12paper}{DR12 paper}  - \citet{2017MNRAS.467.3361W})}, SDSS-III - \citet{2011AJ....142...72E}). We use the largest spectroscopic redshift survey to date to access unprecedented accuracy in the higher-order correlation functions. We explore the growth of structure by measuring redshift evolution of the higher-order correlations, which are expected to grow nonlinearly in time with varying behavior for different orders. 

This paper is organized in a non-standard way. We try to guide the reader through the uncommon quantities by introducing them at the place where we first use them on the data. In this way, there is a more immediate visualization of the mathematical expressions which hopefully promotes understanding. Therefore, we start in \cref{sec:data}~ by describing our usage of the BOSS dataset. \cref{sec:Gg}~ and \cref{sec:MMFs}~ then guide through the  germ-grain model, its application to the data and our transformations of the MFs. \cref{sec:results}~ describes higher-order correlations and relates our analysis of redshift evolution. \cref{sec:SC}~ presents our conclusions. 

\section{BOSS Data}\label{sec:data}

Our sample was collected with the 2.5m Sloan Telescope \citet{2006AJ....131.2332G}. For specifics on photometry and instruments, refer to \citet{1998AJ....116.3040G},\ \citet{1996AJ....111.1748F}, \ \citet{2001ASPC..238..269L}, \ \citet{2002AJ....123.2121S}, \ \citet{2003AJ....125.1559P}, \  \citet{2008ApJ...674.1217P}, \ \citet{2010AJ....139.1628D}, \ as well as the eighth data release \citep{2011AJ....142...72E}. For details on spectroscopic redshift determination see \citet{2013AJ....146...32S} and \citet{2012AJ....144..144B}. Our samples are drawn from the third phase of the SDSS \citep{2000AJ....120.1579Y} Luminous Red Galaxy (LRG) catalog \citep{2001AJ....122.2267E} of the BOSS \citep{2016AJ....151...44D} in DR12 \citep{2015MNRAS.453.1754A}. For specifics regarding the data, see \citet{2016MNRAS.455.1553R}. 

\subsection{Redshift Samples - CMASS and LOWZ}

\begin{table}
\centering
\setlength\tabcolsep{3pt}
\caption{Reduced set of basic parameters of the SDSS DR12 CMASS and LOWZ samples.  The parent samples from \protect\citet{2016MNRAS.455.1553R} yield (their Table 2) 
$\bf{N}_{\mathrm{used}}$ galaxies and the listed effective areas.
We then opt to split the combined sample into two redshift bins, $0.45 \leq z \leq 0.60$ and $0.20 \leq z \leq 0.45$, which we call CMASS and LOWZ, respectively (see also \cref{fig:bins}).  $\bf{N}_{\mathrm{gal}}$ is the number of galaxies in these two redshift slices.
}
\label{tab:param}
\begin{tabular}{p{1.7cm} p{.9cm} p{.9cm} p{.9cm} p{.9cm} p{.9cm} p{.9cm}}
\hline \hline
Sample                              & \multicolumn{3}{c}{CMASS}   & \multicolumn{3}{c}{LOWZ}    \\ \hline
Property                            & NGC     & SGC     & total   & NGC     & SGC     & total   \\ \hline
$\bf{N}_{\mathrm{used}}$             & 568,776 & 208,426 & 777,202 & 248,237 & 113,525 & 361,762 \\
Effective area ($\mathrm{deg}^{2}$) & 6,851   & 2,525   & 9,376   & 5,836   & 2,501   & 8,337  \\
$\bf{N}_{\mathrm{gal}}(\mathrm{our \ bins})$  & 410,617 & 151,003 & 561,620 & 294,091 & 123,719 & 417,810   \\ \hline
\end{tabular}
\end{table}

By using the DR12 dataset, we access an unprecedented number of spectroscopic redshifts. We use a combination of the CMASS and LOWZ samples, described in detail in \citet{2016MNRAS.455.1553R}. The CMASS sample was designed to expand upon the color cuts towards the blue in the SDSS-I and SDSS-II LRG samples, and has a redshift range of $0.40 \lesssim z \lesssim 0.70$. The LOWZ sample was designed to decrease the lower redshift bound of SDSS-I and SDSS-II down to  $z= 0.2$., increasing the effective galaxy count by roughly a factor of 3. Merging those two selections gives the ``complete sample'' which has been used in a number of recent analyses (see \citet{2017MNRAS.470.2617A} and references therein). We consider two redshift bins (\cref{fig:bins}), which we will use CMASS $(0.45 \leq z \leq 0.60)$ and LOWZ $(0.20 \leq z \leq 0.45)$ to refer to from now on. This naming convention means our bins do not correspond to the separation into LOWZ and CMASS used in earlier analyses. The bins we use give a redshift range near triple that used in the \hyperlink{txt:dr12paper}{DR12 paper}. We use both the northern and the southern galactic caps (N- and SGCs) of CMASS and LOWZ for an effective area of $9,376 \ \mathrm{deg}^{2}$. We also make use of the LOWZE2 sample (chunk 2 of the original LOWZ footprint), though it contributes a relatively small amount to the effective area and sample count ($131 \ \mathrm{deg}^{2}$, $2,985$ galaxies). There are a total of $979,430$ galaxies in our two-bin sample, more than double the $410,615$ used in the \hyperlink{txt:dr12paper}{DR12 paper}. The CMASS-South sample is approximately a factor of 2 smaller than its northern counterpart (\cref{tab:param}). This has some effect on the behavior of the MFs, especially at large scales. 

\begin{figure}
\begin{centering}
\includegraphics[width=0.45\textwidth]{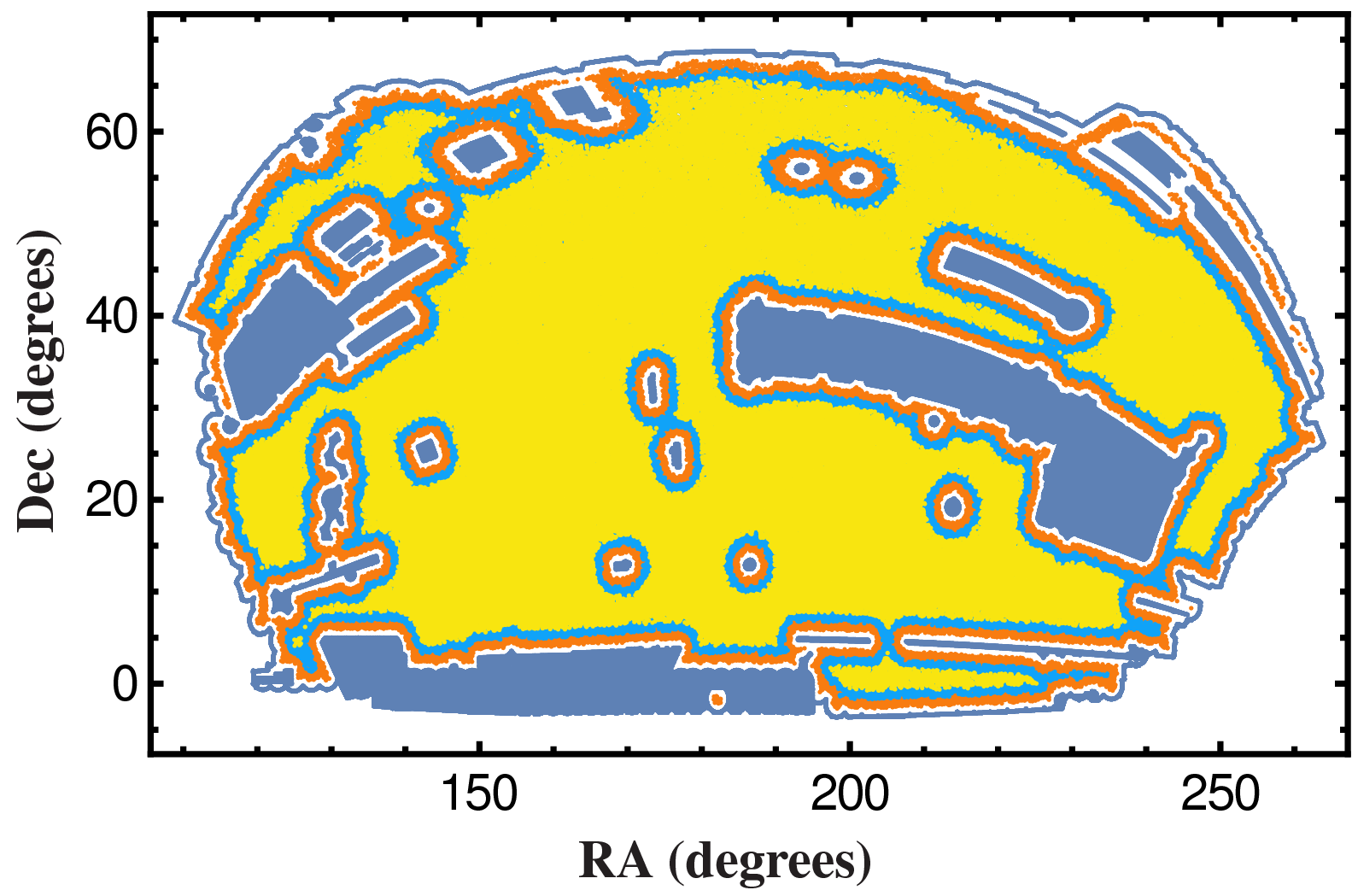}
\includegraphics[width=0.45\textwidth]{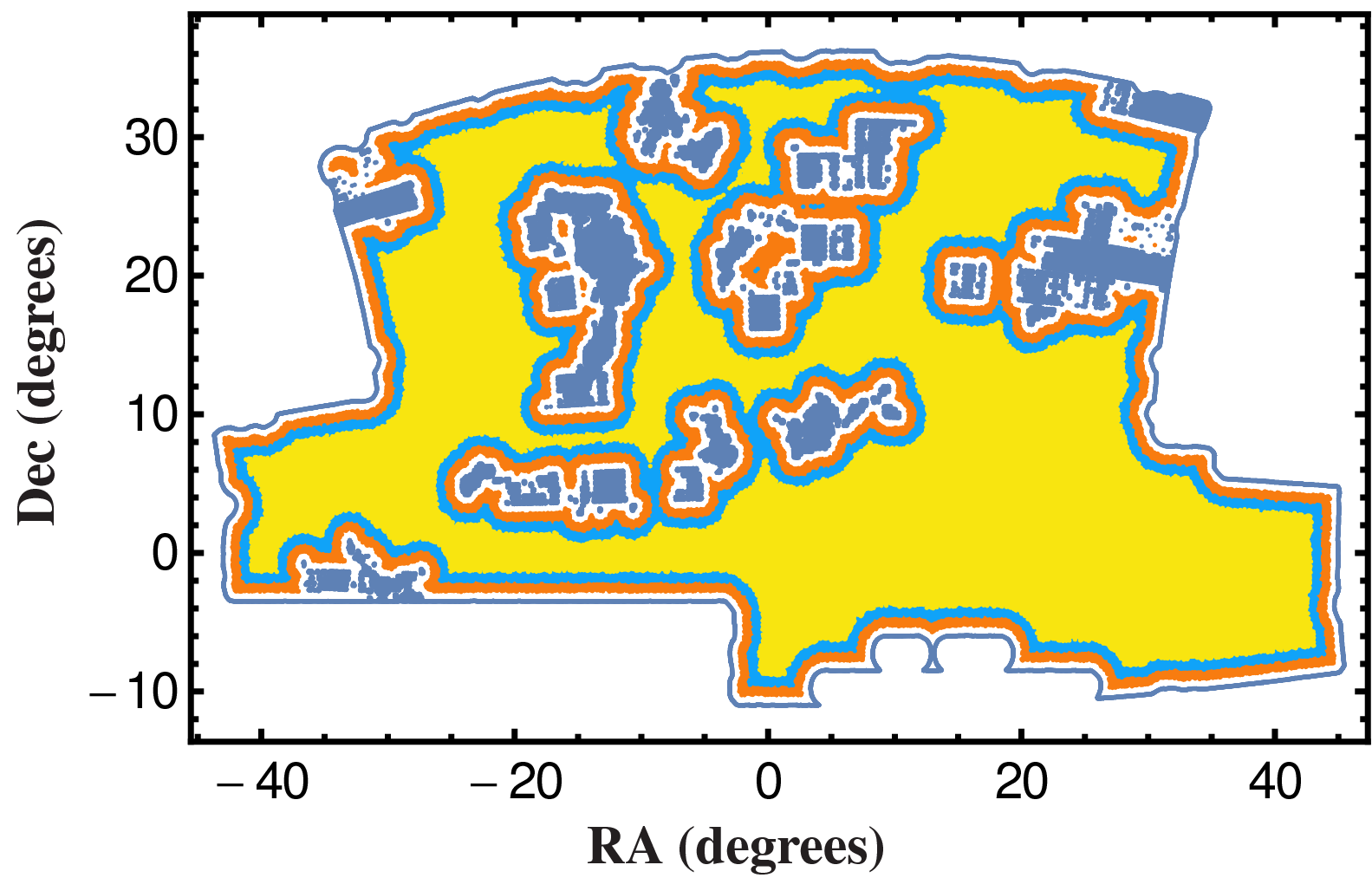}
\end{centering}
\caption{External and internal boundaries for the northern and southern patches analyzed. In blue the regions that we chose to define our boundary. In orange, light blue and yellow a projection is shown of the regions $18$, $36$ and $54 \protect \hMpc$ away from the nearest boundary. The boundary for the northern high-$z$ bin is the same as in \protect\citet{2014MNRAS.443..241W}. The boundary shown here is for LOWZ North excluding the EARLY3 region.}
 \label{fig:mask}
\end{figure}

\subsection{Survey and Corrections}

We used a similar masking process to that of the \hyperlink{txt:dr12paper}{DR12 paper}. In correcting for bad and missing data, we used veto masks to remove galaxies based on effects such as poor spectral seeing, extinction, areas without observations and poor photometric conditions. We applied these veto masks to the mocks in order to make them as similar to the data as possible. Of the remaining galaxies, we considered only those in areas of the survey that were more than a distance of $2R$ from any inner or outer boundary of the data, where $R$ is the scale at which we probe the MFs (see \cref{sec:Gg} for details). This restricted our use of the data, moreso for large values of $R$, but the alternative method of using boundary corrections and random distributions is not effective for MFs.  

The masks for both caps are provided in \cref{fig:mask}. The LOWZ sample especially has a comparatively large number of holes and covers a smaller area of the sky. We see the effects of this on our analysis, especially at low densities. The masks presented in \cref{fig:mask} are a compromise between the need for a contiguous sample and the wish to exclude regions of bad or missing data. We did not check again the extent to which we succeeded in providing an unbiased estimate of the MFs, but the results of the \hyperlink{txt:dr12paper}{DR12 paper} indicate that the periodic box value should be within our error bars. We do some cross check that this is indeed the case in \cref{subsec:growth}.

\subsection{MD Patchy Mocks}
We use mock data realizations to compare the correlations in the BOSS data to those of the cosmological concordance model. The mocks used in our analysis are drawn from the MultiDark(MD)-Patchy mock galaxy catalogs \citep{2016MNRAS.456.4156K}. These mocks were specifically produced to reflect the number density of the data, and were generated by referencing a large N-body simulation (BigMultiDark - \citet{2016MNRAS.457.4340K}). We used the same $\Lambda$CDM cosmology as in the simulation, which the authors called the Planck1 cosmology, with $\Omega_{M} = 0.307115$, $\Omega_{\Lambda} = 0.692885$, $\Omega_{b} = 0.048$, $\sigma_{8} = 0.8288$, and $h=0.6777$. As indicated in the \hyperlink{txt:dr12paper}{DR12 paper}, these mocks are very well suited for comparison when calculating the higher-order parts of the MFs, although there is light tension between the data and the MD-Patchy power spectrum that may be relieved by a $5\%$ reduction in the mock power spectrum amplitude. We used 997 mock survey realizations for CMASS-North, 399 mocks for CMASS-South and LOWZ-North, and 383 mocks for LOWZ-South in our calculations using the CHIPMINK code \citep{2014MNRAS.443..241W}. These large numbers of mock realizations allow us to derive correlation-conscious uncertainties when comparing the data to the concordance model. 

\section{The Germ-grain model}\label{sec:Gg}

In this Section, we provide an abbreviated description of Minkowski Functionals. For a more complete explanation of the germ-grain model, see the \hyperlink{txt:dr12paper}{DR7 paper}. 

\subsection{Minkwoski Functionals}

We apply a result from integral geometry \citep{UBHD152758} and define a set of 4 base functionals for the class of functionals over poly-convex bodies in three-dimensional Euclidean space. For these four Minkowski Functionals, we choose a normalization (in which they are usually denoted $V_{0} - V_{3}$) that relates to familiar geometric quantities as follows:
\begin{equation}
V_{0}=V\;\;;\;\;V_{1}=\frac{S}{6}\;\;;\;\;V_{2}=\frac{H}{3\pi}\;\;;\;\;V_{3}=\chi\;.\label{eq:MinkRelGeo}
\end{equation}

Here $V$ is the volume of the extended body, $S$ is its surface area, $H$ is the integral mean curvature of the surface, and $\chi$ is the topological Euler characteristic or integral Gaussian curvature. To refresh the reader, $H$ is the average of the curvature of a surface over all angles and $\chi$ is the sum of independent components and cavities minus the number of holes.

\begin{figure}
\begin{centering}
\includegraphics[width=0.5\textwidth]{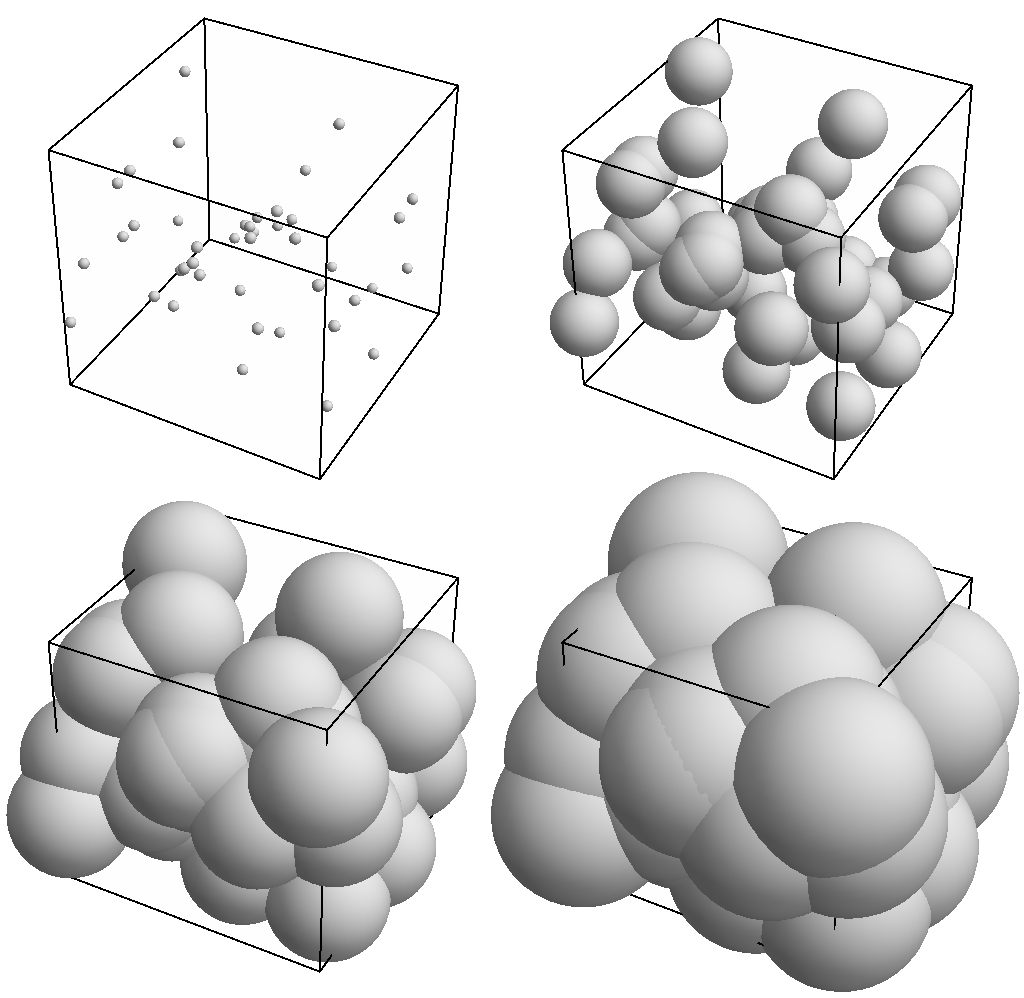}
\end{centering}
\caption{Illustration of the germ-grain procedure to transform a set of galaxies
into an extended body. The galaxy positions 
 are surrounded by balls of a common radius $R$. Then the MFs of the
body formed by the union of all the balls are studied as a function
of $R$. $R$ effectively sets the scale at which correlations are probed. From \protect\citet{2014MNRAS.443..241W}.
 }
\label{fig:boolean}
\end{figure}

The germ-grain model connects MFs and observed galaxy distributions. The germ-grain procedure is to take a point distribution of "germs" and surround each point with a convex body, or "grain." By taking a galaxy distribution with positions $\{\mathbf{x}_{1},\dots,\mathbf{x}_{N}\}$
as the point distribution, and using balls (filled spheres) of radius $R$ as the convex bodies, we create an extended body from the union of the balls. The MFs can now be used to describe the relatively simple extended body that is transformed from the observed galaxy distribution (\cref{fig:boolean}).

The MFs clearly depend on the common ball scaling radius $R$, but they also depend on the sample density $\vr$ of the observed distribution due to the germ-grain construction. The radius $R$ is a tunable parameter that we alter in our analysis to probe different scales of structure. When exploring the density parameter space, we randomly sample from either the mock or data distribution at different values of $\vr$. These values are determined by downsampling at fixed percentages from the height of our bins. At low densities, this tends to erase structure and the MF's approach Poissonity. 

\section{Measuring MFs}\label{sec:MMFs}
In this Section, we describe the quantities we measure for each sample. 
\subsection{Functional Densities}
\begin{figure*}
\begin{center}
\includegraphics[width=3in]{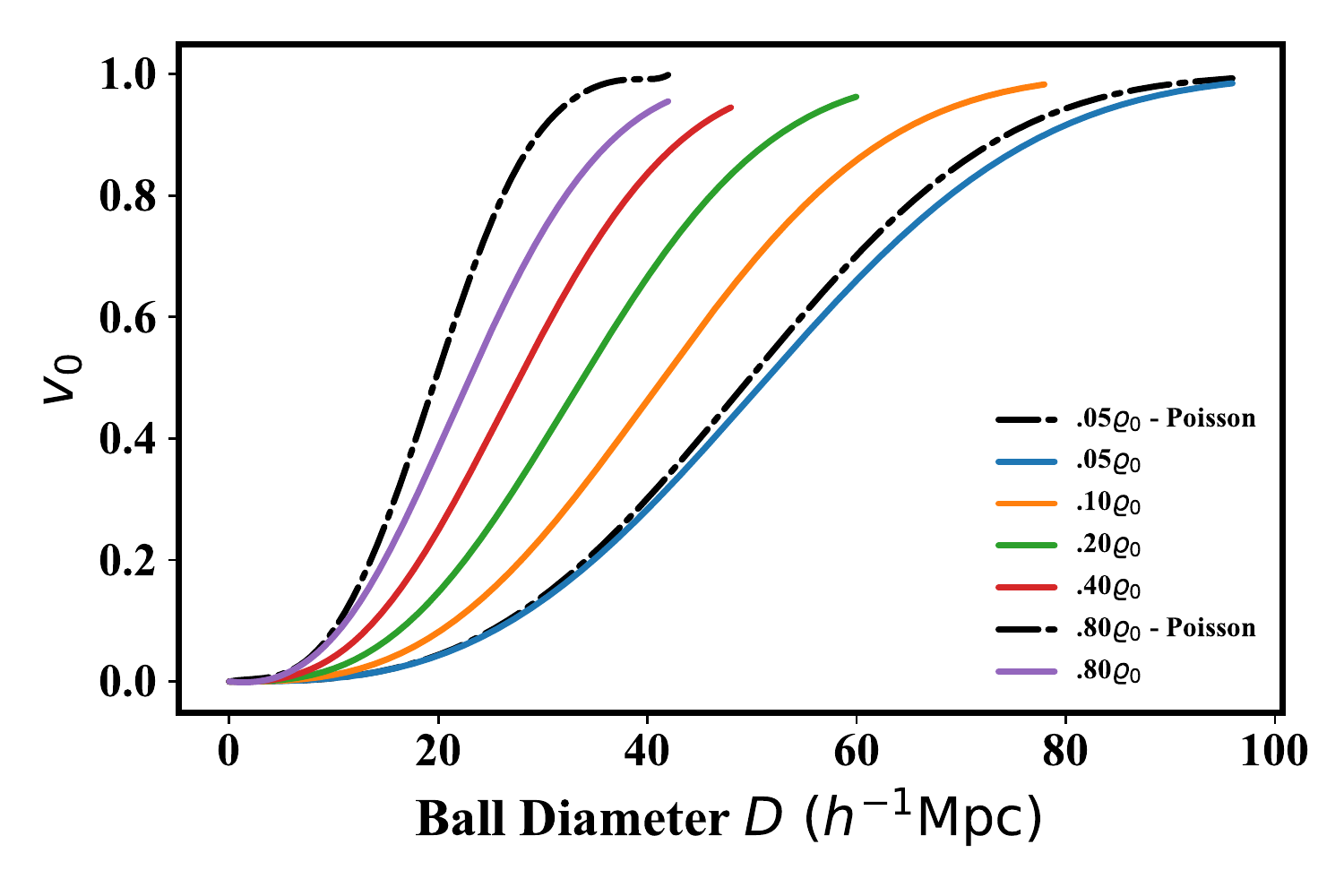}
\includegraphics[width=3in]{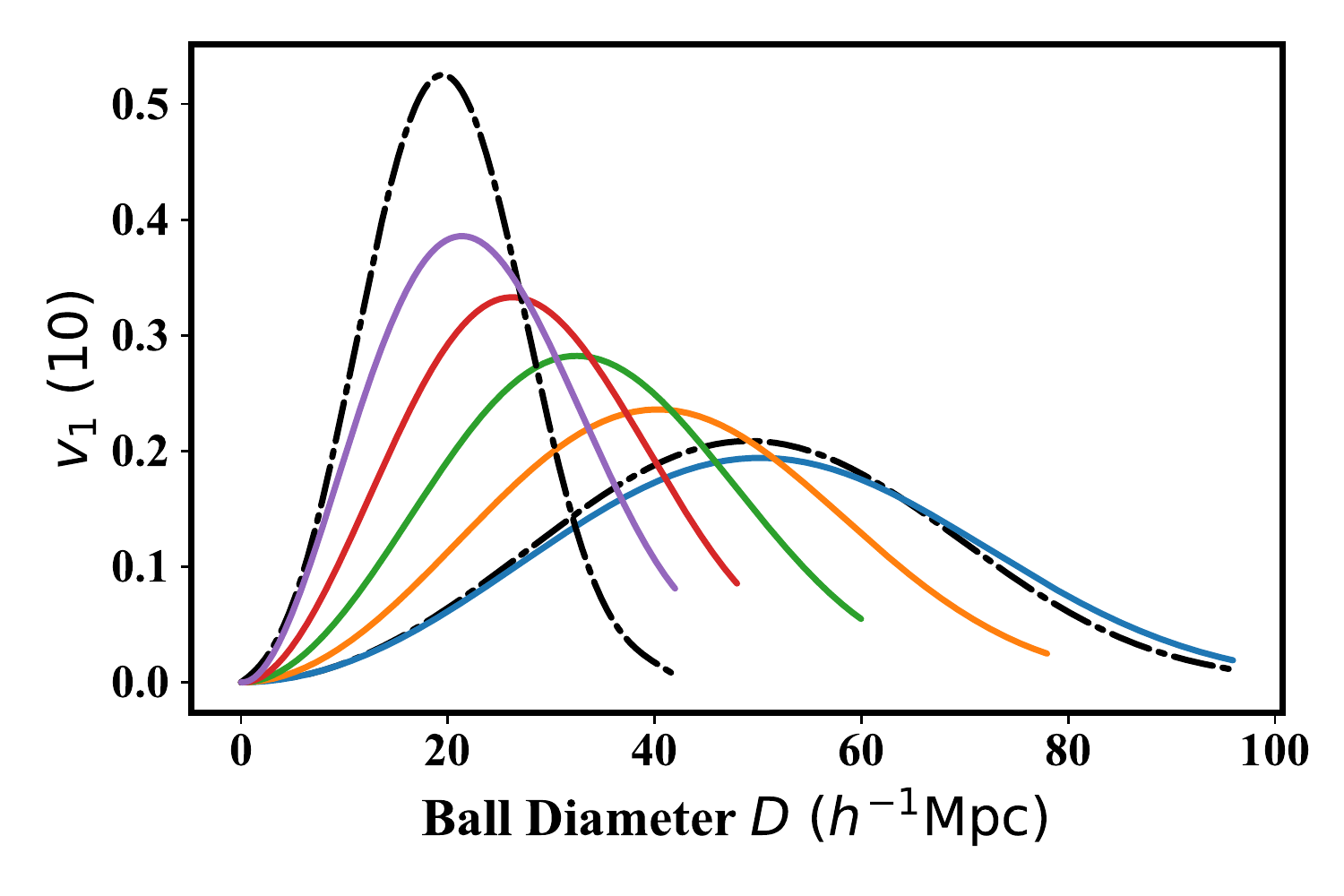}
\includegraphics[width=3in]{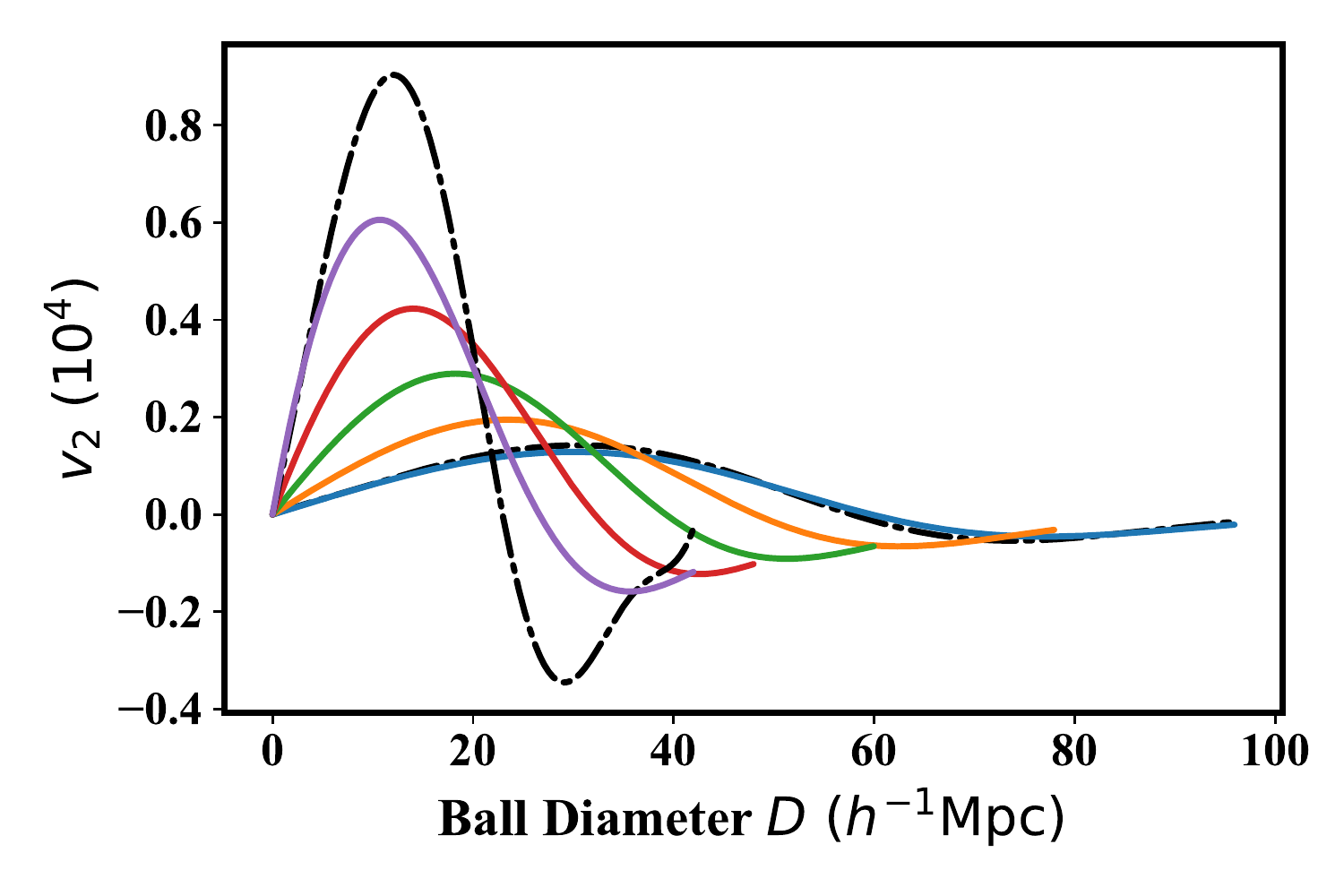}
\includegraphics[width=3in]{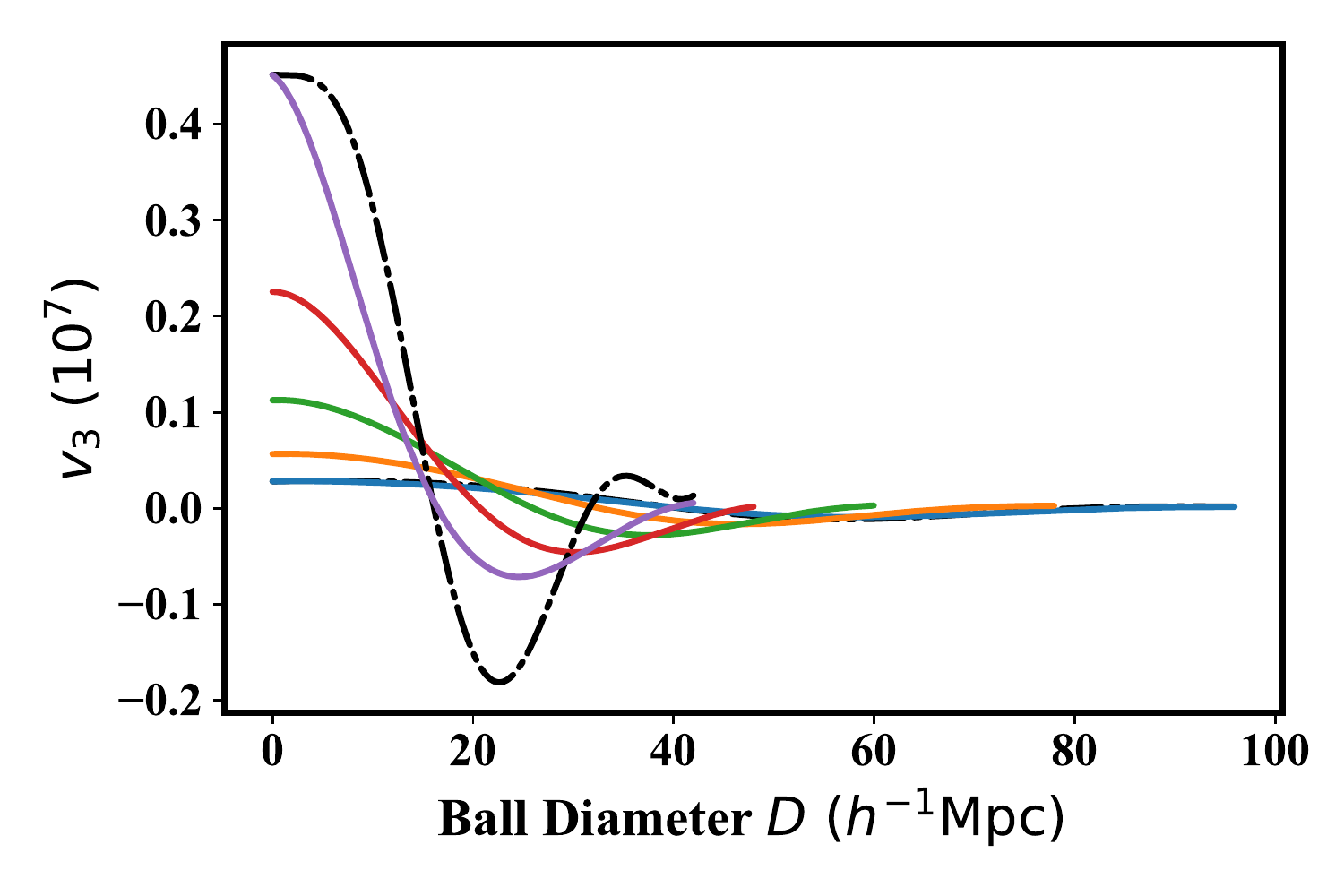}
\caption{MF densities $v_{\mu}$ for the CMASS-North MD-Patchy mock realizations within a sample density range of 5\% to 80\% (in factor of 2 steps) of the reference density $\varrho_{0} = 2.08 \times 10^{-4} h^{3} \mathrm{Mpc}^{-3}$. The most extreme densities are plotted along with their theoretical Poisson point distributions (black dash-dotted lines). At the lowest density the MFs approach the Poisson curves, and at the highest density show the greatest difference from the Poisson curves.}
\end{center}
\label{fig:v_is}
\end{figure*}

To measure the MFs for our samples, we must calculate the MFs of the individual balls. These \emph{partial MFs} are then taken as an ensemble of statistical quantities. Averaging the partial functionals leads to the \emph{global} functionals used in our analysis. Rather than to work with the extrinsic quantities $V_{\mu}$, we use the survey volume $V_{{\rm Survey}}$ to define the intrinsic functional densities $v_{\mu}$:
\begin{equation}
 v_{\mu}\equiv V_{\mu}/V_{{\rm Survey}}\;;\;\mu\in\left\{ 0,\ldots,3\right\}. \label{eq:MinkDens}
\end{equation}
The analytic expressions for these $v_{\mu}$ are given by: 
 \begin{eqnarray}
 \av{v_{0}} & = & 1-e^{-\varrho_{0}\overline{V}_{0}}\;,\nonumber \\
 \av{v_{1}} & = & \varrho_{0}\overline{V}_{1}e^{-\varrho_{0}\overline{V}_{0}}\;,\nonumber \\
 \av{v_{2}} & = & \left(\varrho_{0}\overline{V}_{2}-\frac{3\pi}{8}\varrho_{0}^{2}\overline{V}_{1}^{2}\right)e^{-\varrho_{0}\overline{V}_{0}}\;,\nonumber \\
 \av{v_{3}} & = & \left(\varrho_{0}\overline{V}_{3}-\frac{9}{2}\varrho_{0}^{2}\overline{V}_{1}\overline{V}_{2}+                                                 \frac{9\pi}{16}\varrho_{0}^{3}\overline{V}_{1}^{3}\right)e^{-\varrho_{0}\overline{V}_{0}}\;,\label{eq:MinkDensDef}
 \end{eqnarray}
where $\av{v_{\mu}}$ refer to the average functional densities over the distribution, and $\overline{V}_{\mu}$ refer to the unnormalized modified MFs \citep{2014MNRAS.443..241W}. The $v_{\mu}$ as measured for the mocks are shown in \cref{fig:v_is}. Their behavior as a function of $R$ (or diameter $D$ in this case) and $\vr$ is reasonably intuitive. For $v_0$, as $R$ increases the volume of the survey box occupied by the balls is saturated and $v_0$ approaches unity, and a decrease in density delays this saturation. The other functionals are slightly less transparent, but show that there is some maximum in the uncovered surface area ($v_{1}$) when $R$ is increased, and a similar maximum and minimum in curvature ($v_{2}$) and Euler characteristic ($v_{3}$). For the surface area, a decrease in density will delay the maximum as a function of radius, and for the remaining two MF densities, a decrease in density decreases the amplitude of the functional. These $v_{\mu}$ plainly exhibit non-Poissonity at higher densities, but are relatively opaque to non-Gaussianity, motivating a transformation of these quantities.

\subsection{Transforming the Densities}

Observed galaxy distributions have significant structure, and thus must employ the full form of $\overline{V}_{\mu}$ in \cref{eq:MinkDensDef}. But, for the case of a Poisson distribution, the unnormalized modified MFs $\overline{V}_{\mu}$ are given by the more familiar MFs of a ball $B$, i.e. $\overline{V}_{\mu}=V_{\mu}\left(B\right)$ with
\begin{eqnarray}
&  & V_{0}\left(B\right)=\frac{4\pi}{3}R^{3}\;\;;\;\;V_{1}\left(B\right)=\frac{2}{3}\pi R^{2}\;\;;\nonumber \\
&  & V_{2}\left(B\right)=\frac{4}{3}R\;\;;\;\;V_{3}\left(B\right)=1\;.\label{eq:MinkBall}
\end{eqnarray}
These are natural and convenient to work with, and we use the Poisson case as a reference in analyzing distributions with more structure. Following from this choice, we adopt the dimensionless transformed MFs $\eta_{\mu}$: 
\begin{equation}
\eta_{\mu}\equiv\overline{V}_{\mu}/V_{\mu}\left(B\right)\;.\label{eq:etaDef}
\end{equation}

\begin{figure*}
\begin{center}
\includegraphics[width=3in]{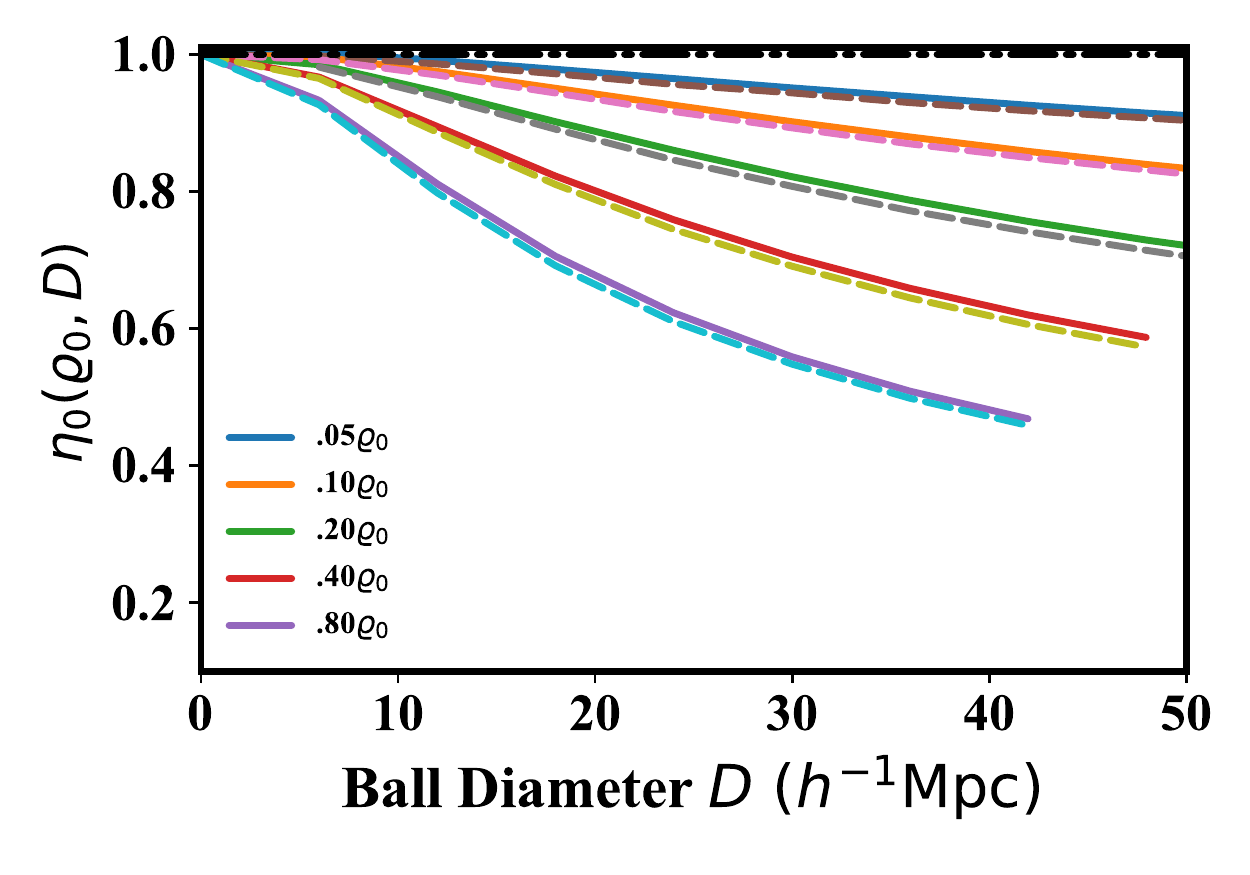}
\includegraphics[width=3in]{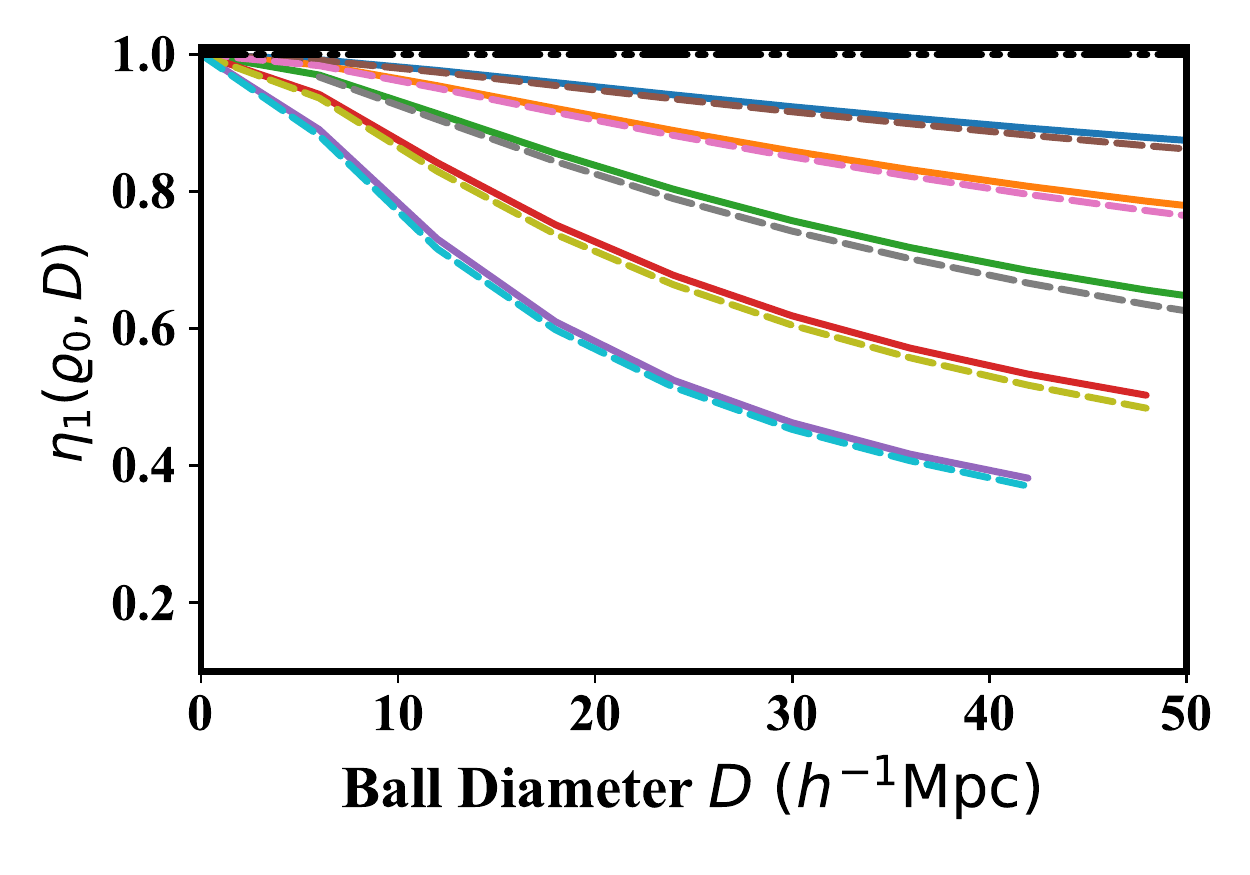}
\includegraphics[width=3in]{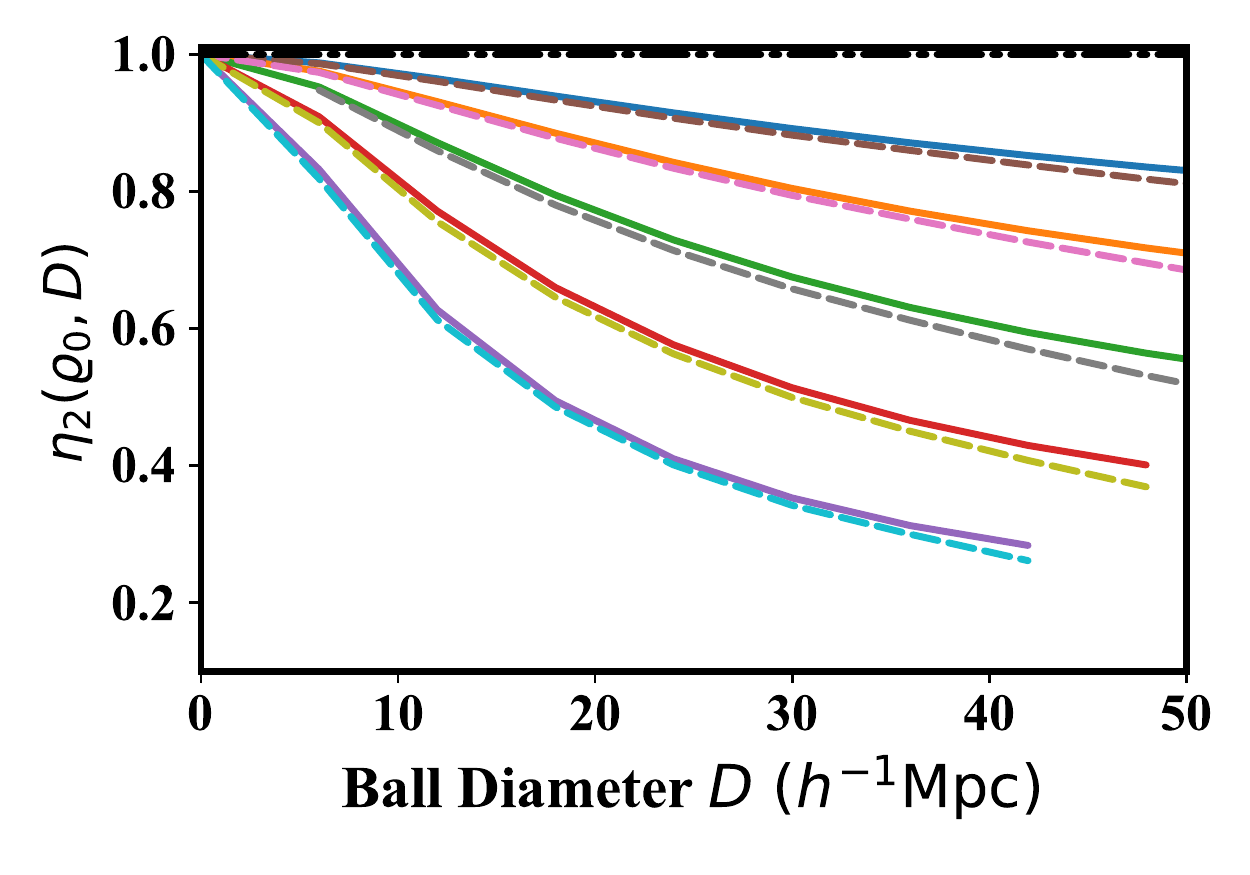}
\includegraphics[width=3in]{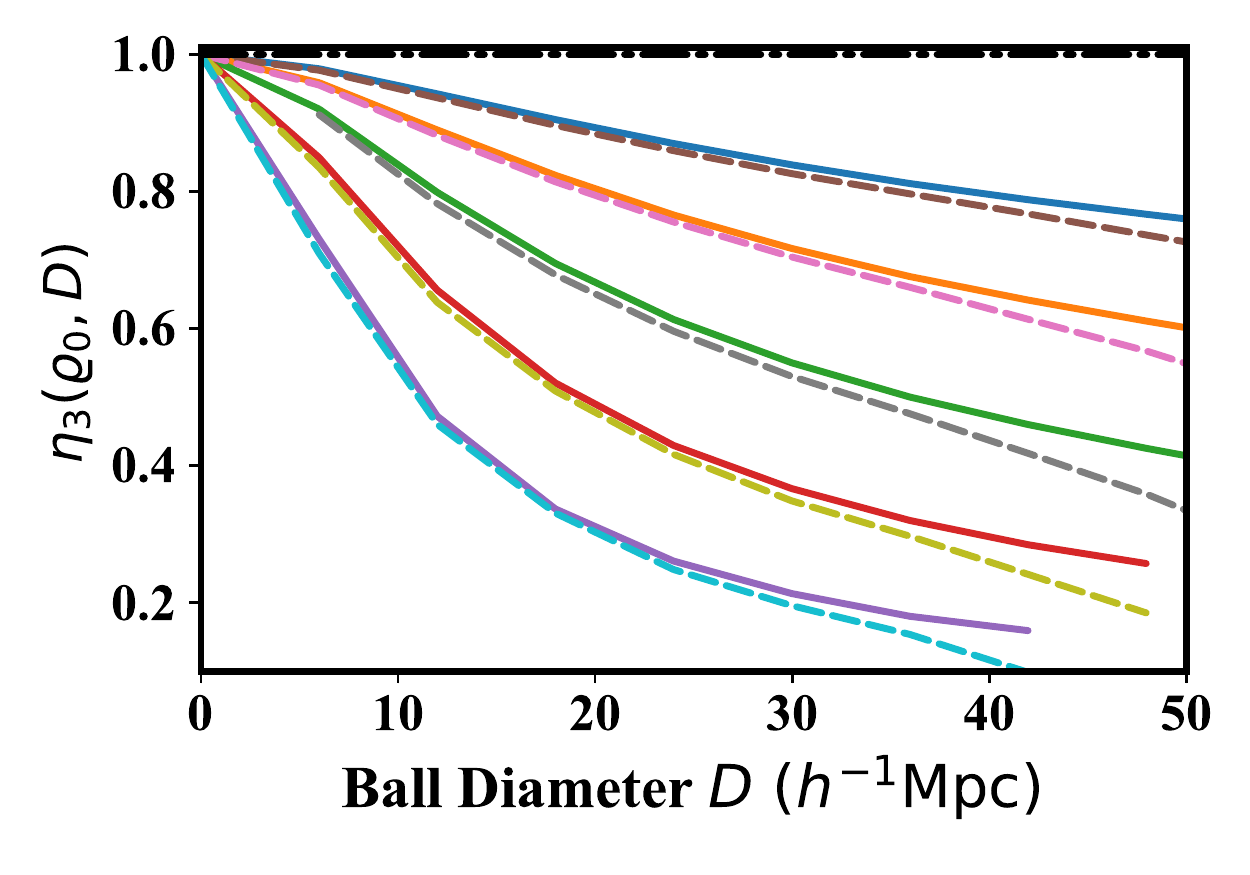}
\end{center}
\caption{Dimensionless transformed MFs in the Northern Galactic Cap evaluated for 997 (CMASS) and 399 (LOWZ) SDSS DR12 MD-Patchy mocks plotted as a function of ball diameter $D$ for densities 5\% to 80\% of the reference density $\varrho_{0} = 2.08 \times 10^{-4} h^{3} \mathrm{Mpc}^{-3}$. CMASS-North values are solid, LOWZ-North values are dashed. These quantities can be more directly related to the correlation functions than the $v_{\mu}$ can. Significant difference from the Poisson case is shown by deviation in the MFs from the black dash-dotted line at 1. The difference in the functionals between the redshift bins is visible to a slight degree across all the MFs.}
\label{fig:etas}
\end{figure*}

The mapping from $v_{\mu}$ to $\eta_{\mu}$ retains the information about the MFs as a function of $R$ and $\vr$, but the latter provides a better perspective in terms of the higher-order correlations. To see this, we can rewrite the dimensionless MFs as the following power series:
\begin{equation}
\eta_{\mu}=\sum_{n=0}^{\infty}\frac{c_{\mu,n+1}}{\left(n+1\right)!}\left(-\varrho_{0}V_{0}\left(B\right)\right)^{n}
\label{eq:PowerSeriesDecomp}
\end{equation}
where $c_{\mu,1} = 1$ and 
\begin{eqnarray}
c_{\mu,n+1}\left(R\right) & = & V_{0}^{-n}\left(B\right)\int\xi_{n+1}\left(0,\bx_{1},\dots\bx_{n}\right)\times\label{eq:MinkIntegr}\\
 &  & \frac{V_{\mu}\left(B\cap B_{\bx_{1}}\cap\dots\cap B_{\bx_{n}}\right)}{V_{\mu}\left(B\right)}\d^{3}x_{1}\dots\d^{3}x_{n}\nonumber 
\label{eq:cexp}
\end{eqnarray}
where the weights $V_{\mu}\left(B\cap B_{\bx_{1}}\cap\dots\cap B_{\bx_{n}}\right)/V_{\mu}\left(B\right)$ in the integrand may be derived by calculating the MFs of the intersection of $n$ balls. 

To emphasize the structure: Each term of the series \cref{eq:cexp} is the integral of the product of the $n$-point correlation function $\xi_{n+1}$ and a geometric weighting function that only depends on the balls (and therefore $R$). The factor that includes the sample density $\vr$ dependence is independent of the correlation-dependent integral. Now we have a quantity conveniently defined on the unit interval $[0,1]$ (for standard cosmological structure) that can be more easily compared in terms of our parameters. 

\cref{fig:etas} shows representative density samples (5\% - 80\% of $\vr = 2.08 \times 10^{-4} \ h^{3} \ \mathrm{Mpc}^{-3}$) of $\eta_{\mu}$ as a function of $D$ and gives a clearer measure of the non-Poissonian behavior of the MFs. The MFs deviate further from Poisson behavior as $D$ and $\vr$ increase, approaching the structure of the full galaxy distribution, and illustrating the increased non-Poissonity at large scales. The MFs are on the order of $1\%$ different between the redshift bins, so already the redshift evolution in the MFs is evident. However, we have yet to isolate higher-order contributions of the MFs, which we discuss in the next section.

\begin{figure*}[]
\begin{center}
\includegraphics[width=3in]{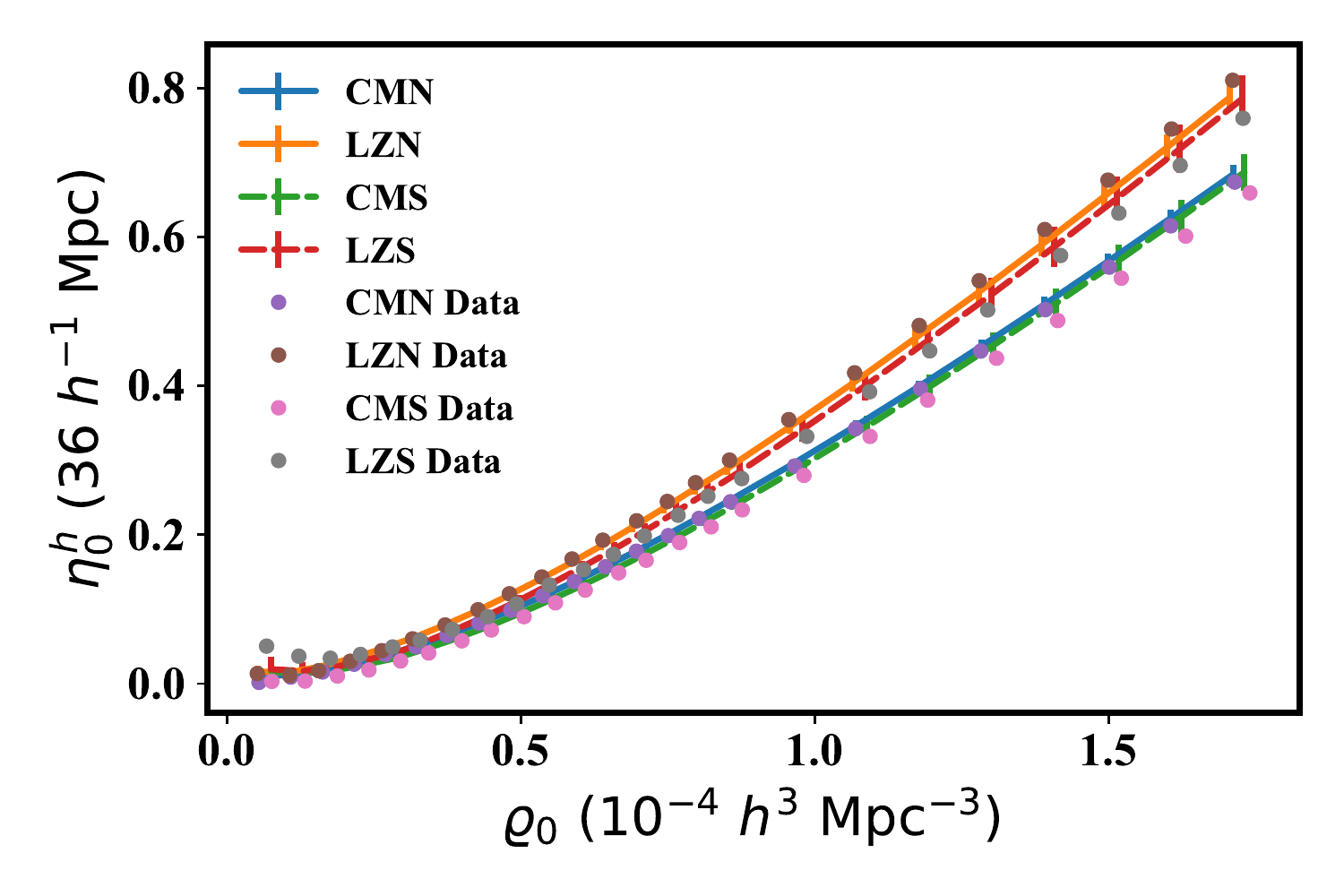}
\includegraphics[width=3in]{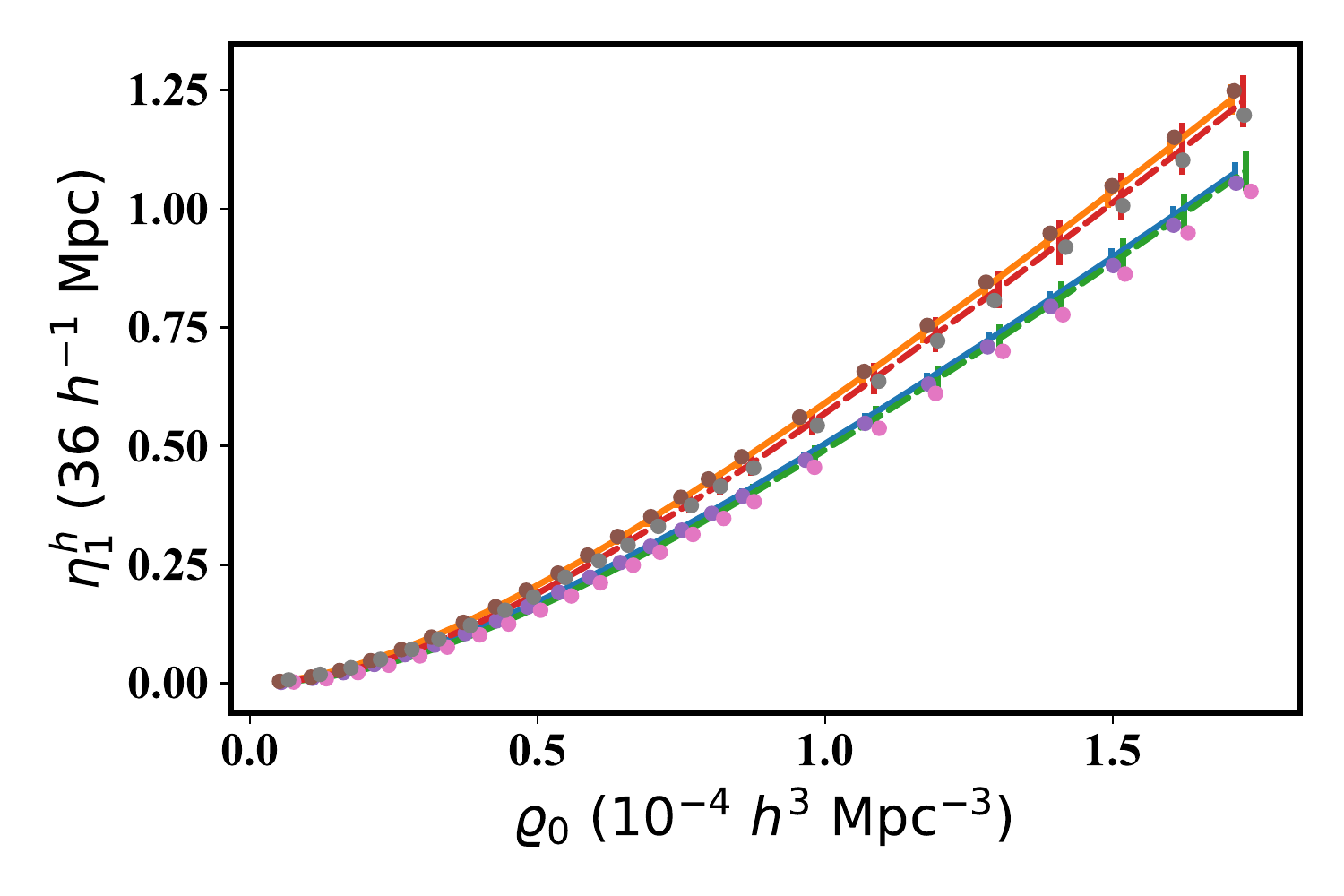}
\includegraphics[width=3in]{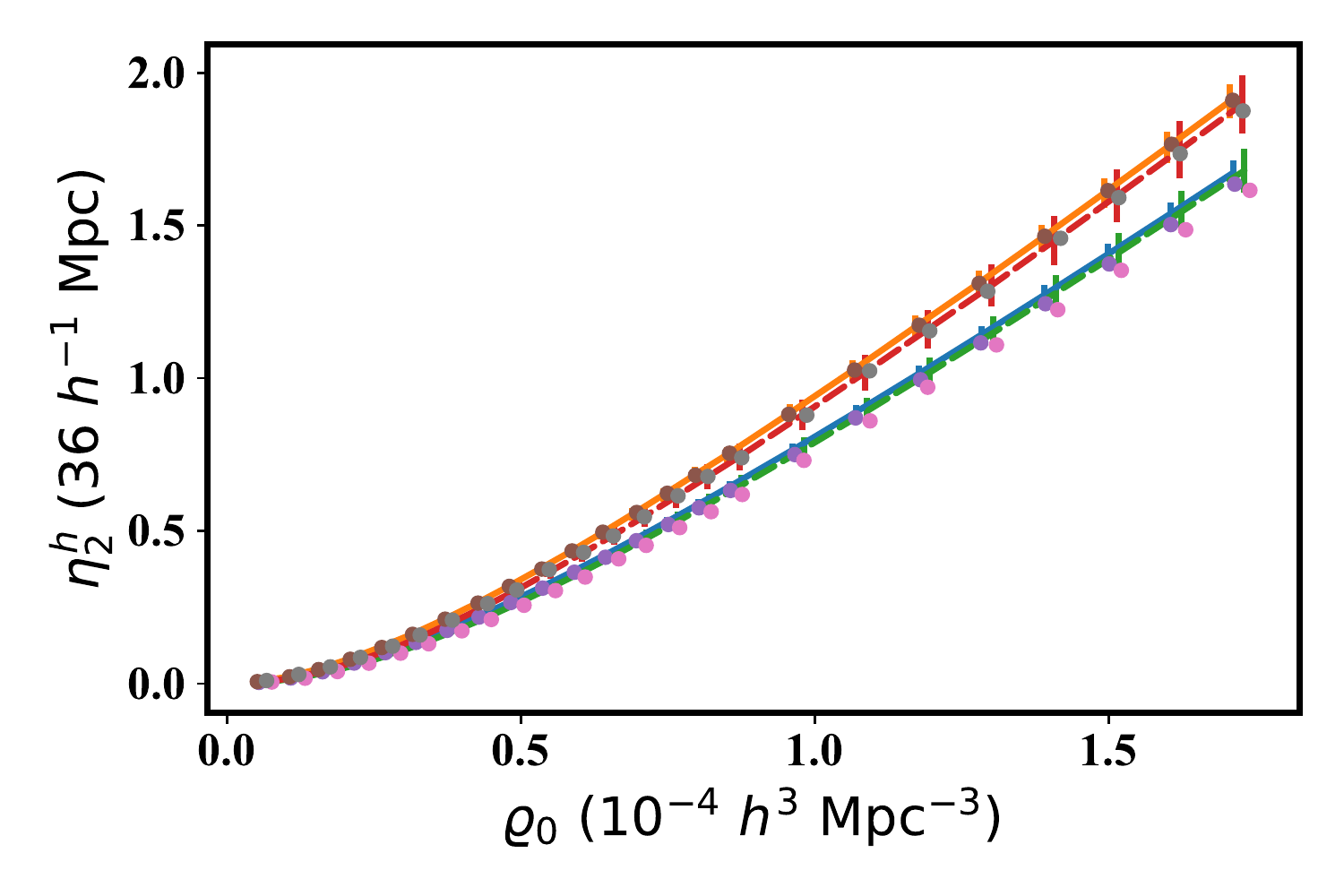}
\includegraphics[width=3in]{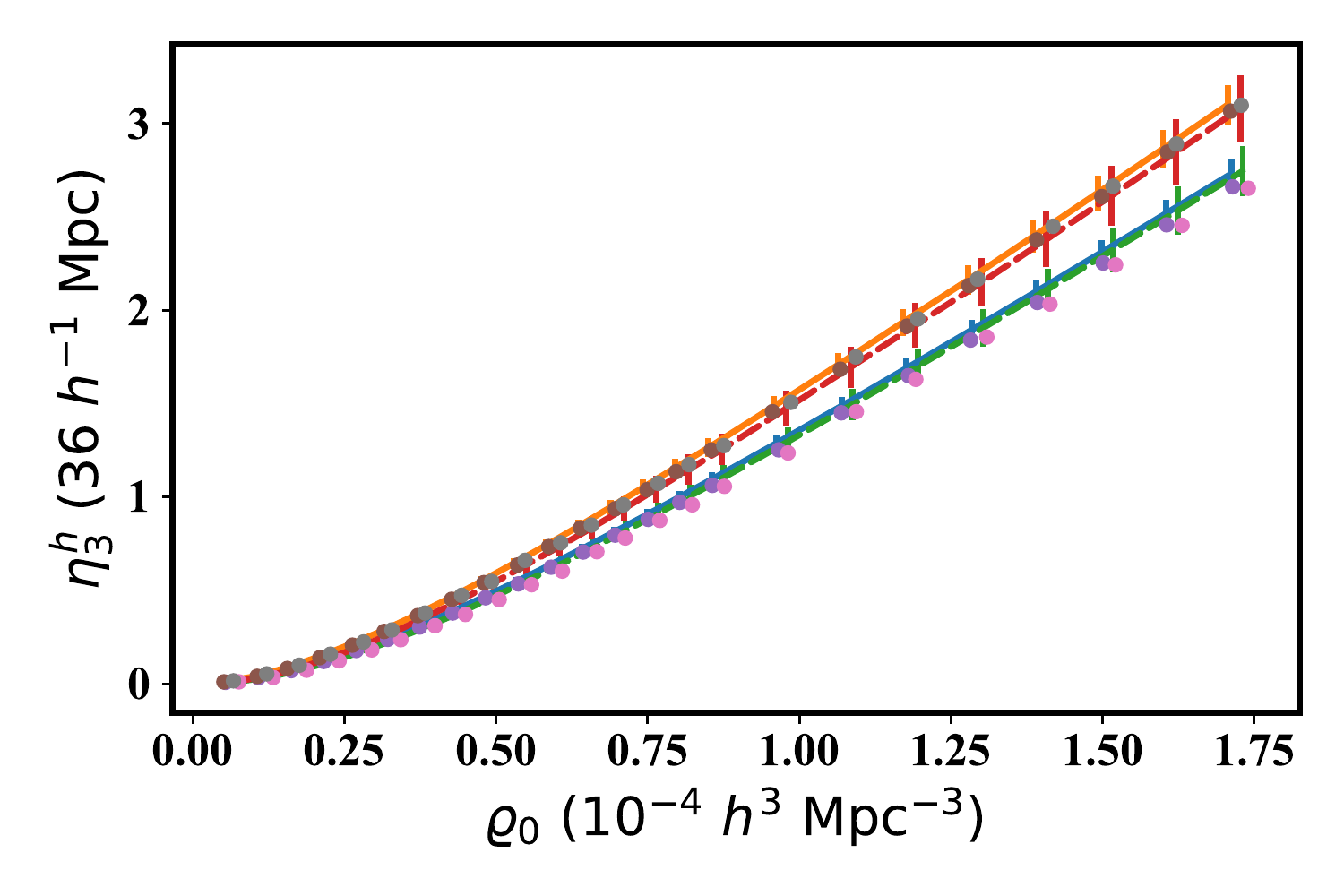}
\end{center}
\caption{The higher-order contribution to the MFs $\eta_{\mu}^{h}$ as a function of density $\vr$. The contribution of the known two point correlations has been removed by subtracting \cref{eq:MinkGaussExp} from \cref{eq:PowerSeriesDecomp}. The lines are the MD-Patchy mock values with $1-\sigma$ error bars on the mean. The dots are the values derived from the data.  Already on this figure we can see a clear split between the two redshift bins (see also \cref{tab:chisq_growth}). Moreover, both the different regions on the sky and the data and the mocks agree reasonably well (see also \cref{tab:chisq_sky}).}
\label{fig:etah_data}
\end{figure*}

\section{Higher-order Contributions to the MFs}\label{sec:results}
This Section describes our analysis of the isolated higher-order contributions to the MFs. We start with the definition of the higher-order quantity we analyze and then turn to the main results of the paper, i.e. the growth of these higher-order contributions.

\subsection{The Two-point Contribution}

To measure higher-order correlations, we subtract the two-point correlation function contribution to the MFs. To do this, we must first calculate the two-point term. Ignoring the higher-order terms ($n\geq 3$) of \cref{eq:PowerSeriesDecomp}
we are left with:
\begin{equation}
\eta_{\mu}=1-\frac{\varrho_{0}}{2}\int\d^{3}x_{1}\xi_{2}\left(0,\bx_{1}\right)\frac{V_{\mu}\left(B\cap B_{\bx_{1}}\right)}{V_{\mu}\left(B\right)}\;.
\label{eq:MinkGaussExp}
\end{equation}

We assume isotropic correlations, and upon transforming to spherical coordinates \cref{eq:MinkGaussExp} becomes:
\begin{equation}
\eta_{\mu}=1-2\pi\varrho_{0}\intop_{0}^{2R}\frac{V_{\mu}\left(R,r\right)}{V_{\mu}\left(B\left(R\right)\right)}\xi_{2}\left(r\right)r^{2}\d r\;,
\label{eq:MinkGaussExpb}
\end{equation}
where the integral is zero if the centers of the balls are more than $2R$ apart. This is the same as \cref{eq:PowerSeriesDecomp} for $n=1$ barring the factor of $V_0(B)$.

We can write the weight functions $V_{\mu}\left(B\cap B_{\bx_{1}}\right)$ as:
\begin{eqnarray}
V_{0}\left(R,r\right) & = & \frac{1}{12}\pi\left(2R-r\right)^{2}\left(r+4R\right)\;\;;\label{eq:V0r}\\
V_{1}\left(R,r\right) & = & \frac{1}{3}\pi R\left(2R-r\right)\;\;;\\
V_{2}\left(R,r\right) & = & \frac{2}{3}\left(2R-r\right)+\frac{2}{3}R\sqrt{1-                                                   \left(\frac{r}{2R}\right)^{2}}\arcsin\left(\frac{r}{2R}\right);\label{eq:Mink2TwoBall}\\
V_{3}\left(R,r\right) & = & 1\;\;,\label{eq:V3r}
\label{eq:weights}
\end{eqnarray}
where $R$ is again the radius of the balls and $r$ is the integration variable. The weights then conveniently define integration windows for $\xi_{2}$ for the different functionals, and as $\mu$ increases, so does the effective scale at which the structure is probed (see the \hyperlink{txt:dr12paper}{DR12 paper}  for a more detailed discussion). 

\begin{table}
\centering
\caption{Significance of the detection of higher-order clustering in the MFs evaluated at $D = \mathrm{36 \ h^{-1} \ Mpc}$. The $\chi^{2}$ values quantify the significance of the deviation of the higher-order functionals $\eta_{\mu}^h$ of the data from the zero line in \cref{fig:etah_data}.  N refers to North, S to South, and CMASS and LOWZ to our bins of \cref{fig:bins}. There are $24$ degrees of freedom, and hence we find a highly significant contribution of the higher-order part of the MFs for all analyzed samples.  The decrease with $\mu$ is due to the increasing error bars induced by the inversion of \cref{eq:MinkDensDef}. The decrease from left to right reflects the relative sizes of the samples. Having this level of significance is important for our study of the changes of theses quantities with redshift.
}
\label{tab:chisq_det}
\begin{tabular}{lcccc}
\hline
\hline
$\mu$ &CMASS-N & LOWZ-N & CMASS-S & LOWZ-S\\
\hline
0 & 5385 & 3431 & 1796 & 1267 \\
1 & 5067 & 3098 & 1666 & 1096 \\
2 & 3993 & 2463 & 1230 & 851 \\
3 & 2664 & 1693  & 794 & 601 \\
\hline
\end{tabular}
\end{table}

\begin{figure*}
\begin{center}
\includegraphics[width=3.1in]{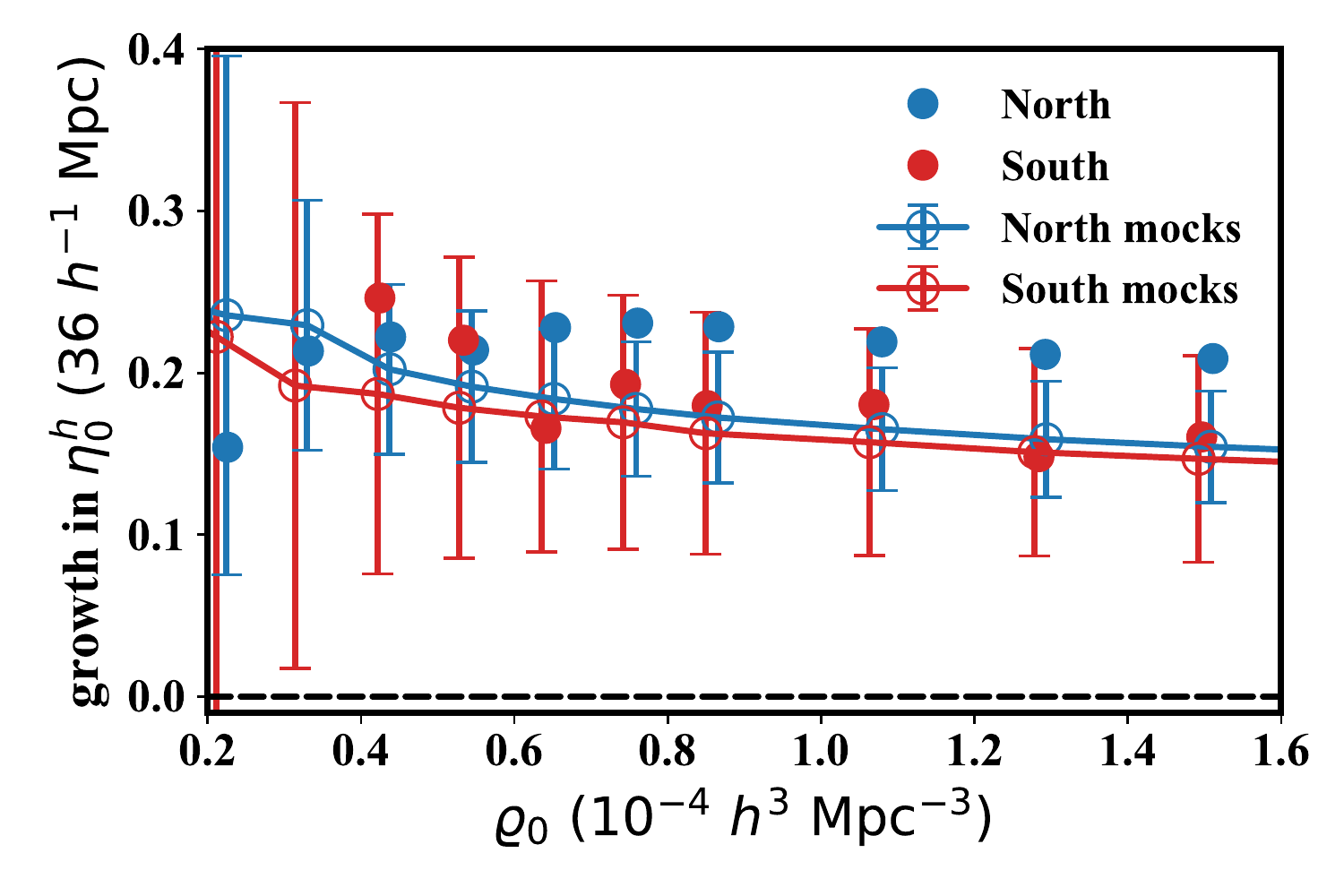}
\includegraphics[width=3.1in]{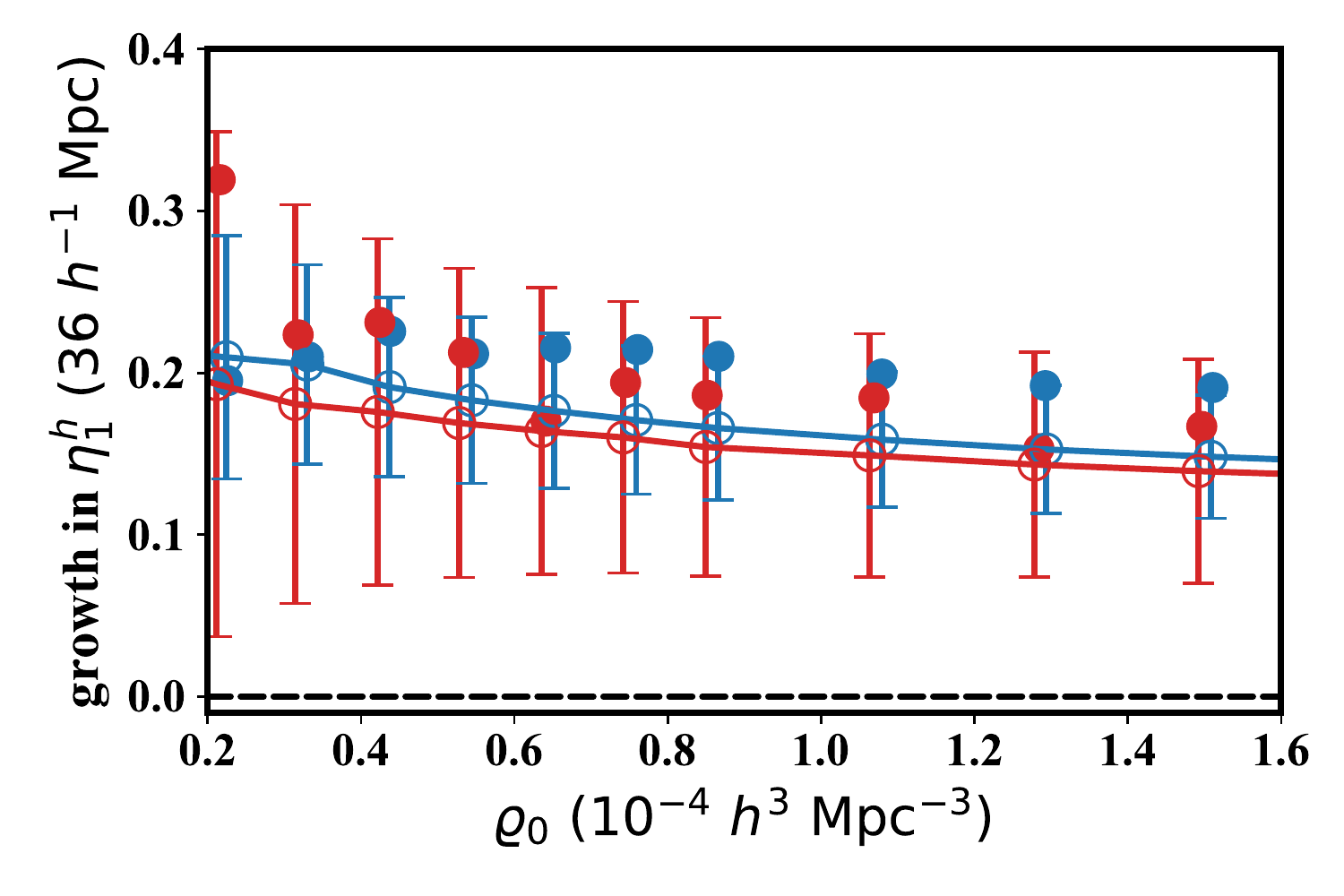}
\includegraphics[width=3.1in]{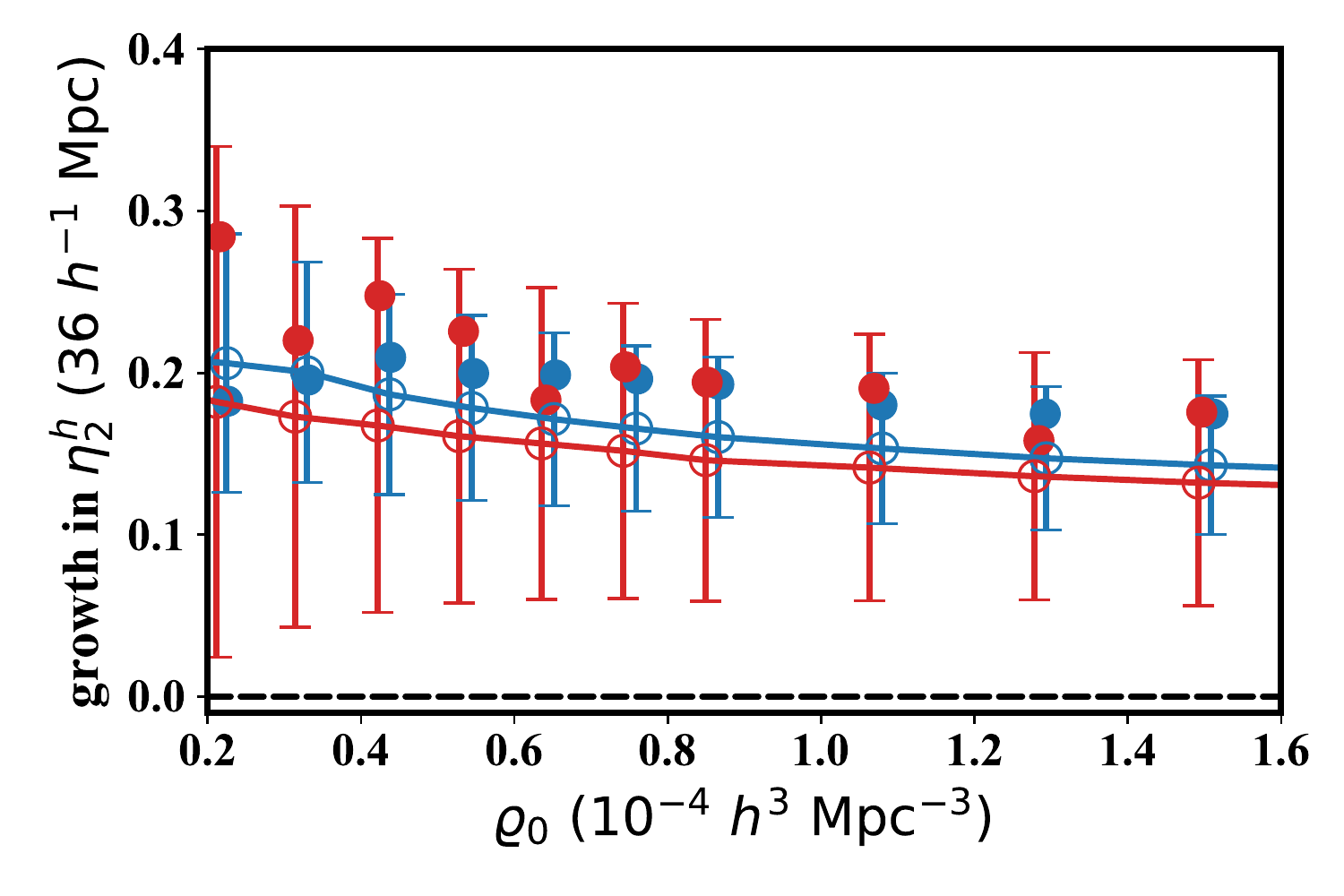}
\includegraphics[width=3.1in]{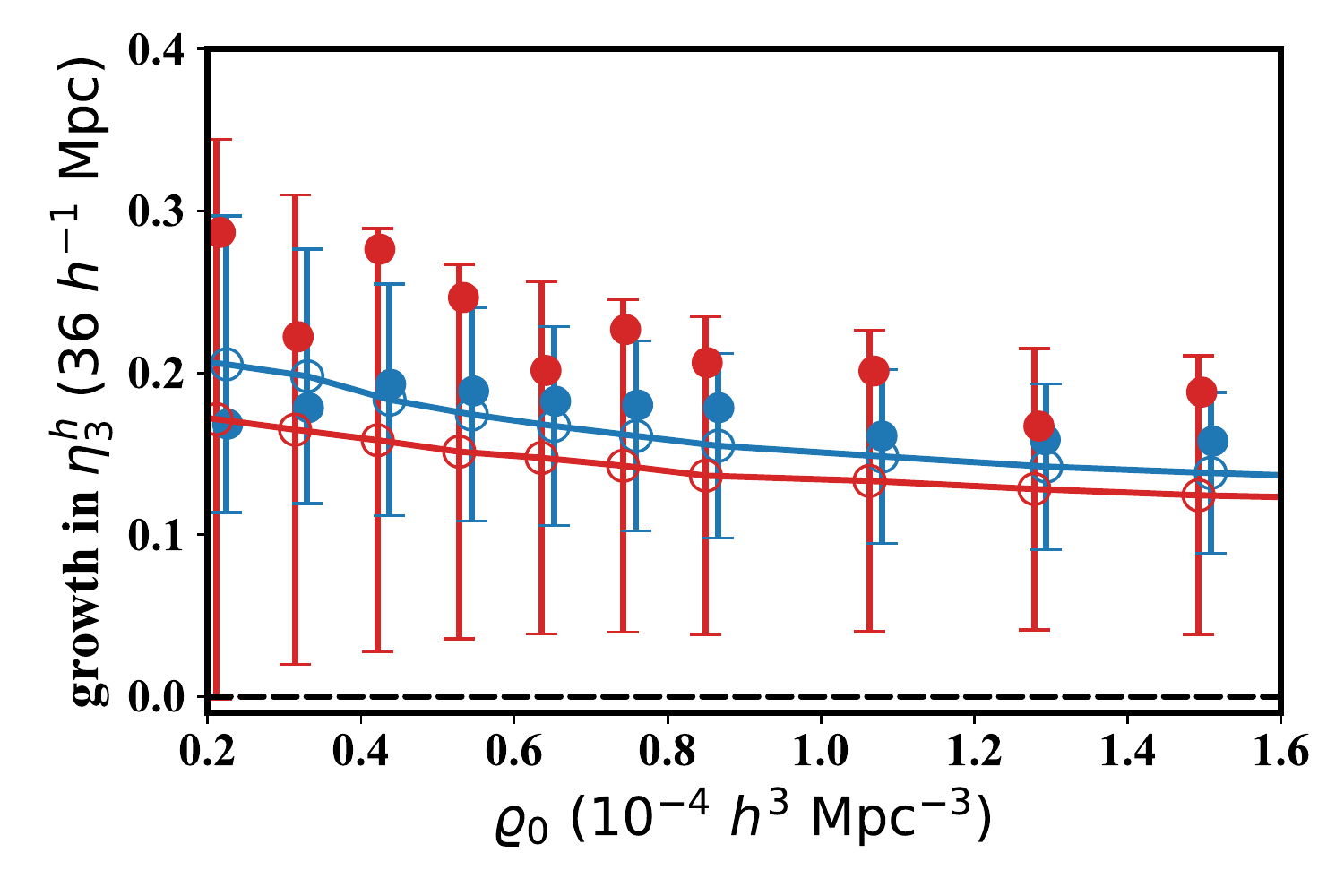}
\end{center}
\caption{The percentage growth of the higher-order term of the functionals $\eta_{\mu}^{h}$ between redshift bins. We find approximately $15\%-25\%$ growth from the higher redshift bin to the lower redshift bin. The MD-Patchy mock values are shown by the solid lines with hollow points and $1-\sigma$ error bars for one mock realization each, and the data by the solid points. The zero line is shown in black for comparison to the no-growth case.
}
\label{fig:ehper}
\end{figure*}

\subsection{Higher-order Functionals and Growth}\label{subsec:growth}
We isolate the higher-order contributions to the MFs to study the higher order correlations of the observed galaxy distribution. Subtracting the two-point contribution \cref{eq:MinkGaussExpb} from the power series expression for $\eta_{\mu}$, we take only the terms with $n \geq 2$ in \cref{eq:PowerSeriesDecomp} as the higher-order part of the functionals, which we define as $\eta_{\mu}^{h}$. The two-point contribution is obtained from a direct measurement of the two-point correlation function monopole by performing the integral in \cref{eq:MinkGaussExpb}.

The higher-order functionals $\eta_{\mu}^{h}$ are a significant portion of the calculated functionals $\eta_{\mu}$. \cref{fig:etah_data} shows $\eta_{\mu}^{h} (D = 36  \ h^{-1} \ \mathrm{Mpc})$ for both redshift samples and both hemispheres. We measure the significance of the detection of higher-order clustering by computing $\chi^{2}$ values with 24 degrees of freedom for the mocks and the zero line in \cref{tab:chisq_det} and find values of order $10^{3}$, indicating a highly significant contribution of higher-order correlations to the structure of the distribution. All $\chi^{2}$ values are computed using the correction to the estimated inverse covariance matrix of \citet{2014MNRAS.439.2531P}. The clear detection of the higher-order contributions to the MFs is important for this paper's goal of studying the redshift evolution of higher-order clustering. \cref{tab:chisq_det} also illustrates the enormous increase in our ability to probe higher-order correlations via $\eta_{\mu}^{h}$ due to the large increase in sample size and volume relative to DR7.

In the limit of low sampling density in \cref{fig:etah_data}, the higher-order part seems negligible, reflecting the near two-point structure characteristic of a completely Gaussian point distribution. Though it seems intuitive to expect this in the low-density limit, there is no theorem that would require a structured point process to pass through a stage of Gaussianity when the density is reduced. Regardless, the two-point behavior increasingly fails to describe the higher-order MFs as the sample density increases, meaning the integrated three and higher point functions must be non-zero, quantitatively capturing the non-Gaussianity of the density field. 

\begin{table}
\centering
\caption{Comparison of the NGC and SGC in each redshift bin. The $\chi^{2}$ values of significance are for the deviation in the MFs of the data in the NGC and SGC. We attempt to correct for potential systematics by subtracting the separation of the mean of 399 mocks from the difference in the data. $\chi^{2}$ values are for $\eta_{\mu}^{h}$ in each redshift bin at $D = \mathrm{36 \ h^{-1} \ Mpc}$ for 24 degrees of freedom. 
}
\label{tab:chisq_sky}
\begin{tabular}{lcccc}
\hline \hline
$\mu$ & \multicolumn{2}{c}{CMASS} & \multicolumn{2}{c}{LOWZ}\\
\hline
 & $\chi^{2}$ & $p$-value & $\chi^{2}$ & $p$-value \\
\hline
0 & 14.1 & 0.055 & 33.3 & 0.098 \\
1 & 18.3 & 0.211 & 40.2 & 0.020 \\
2 & 19.2 & 0.257 & 39.4 & 0.025 \\
3 & 21.3 & 0.378 & 32.2 & 0.122 \\
\hline
\end{tabular}
\end{table}

The data from the northern and southern sky in \cref{fig:etah_data} shows consistent higher-order clustering. \cref{tab:chisq_sky} quantifies this by computing the difference between the $\eta_{\mu}^{h}$ of the data in the south and the north for each of our two redshift bins.  Because the analysis of the mocks reveal a small shift between the hemispheres due to the different boundary corrections, we apply the difference between the mean of the mock results as a correction to the hemispheric residual of the data before computing $\chi^2$.
As we are testing a residual difference, we use the sum of the south and north covariance matrices to estimate the significance of this residual.  We use the 24 density points in \cref{fig:etah_data} and we evaluate both, mocks and data, at common values for the density. We find that the residuals have $\chi^2$ values ranging from 14.1 to 40.2. Given that we have 24 degrees of freedom, these are unsurprising values, with no case below 2\% probability.  Therefore, we conclude that the results in the SGC and NGC are statistically consistent with each other.

\begin{figure}
\begin{center}
\includegraphics[width=3.1in]{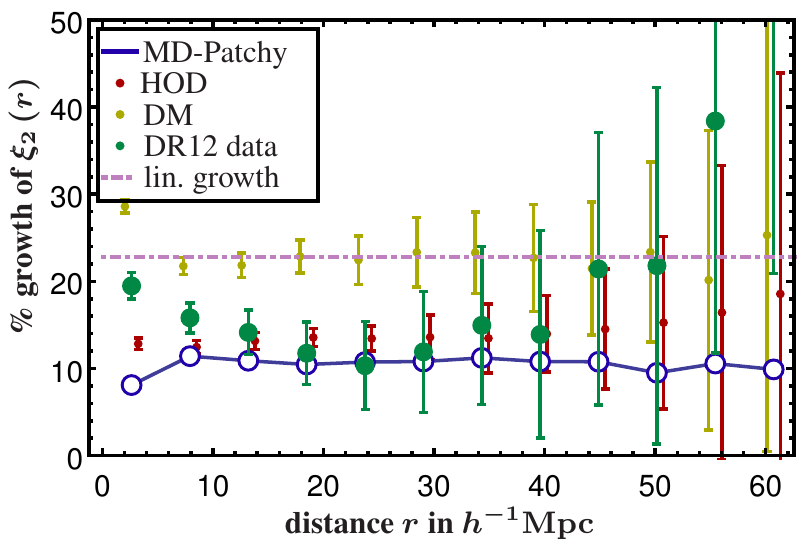}
\end{center}
\caption{Comparison of the growth of two-point correlations between our two redshift bins for different cases. We use only data from the NGC here. Blue hollow points joined by the line show the mean of the MD-Patchy mocks. The red points are for the galaxy samples derived from our own HOD model, and the yellow points show the growth of the underlying dark matter field. The green solid points are the values for the galaxies from the final DR 12 sample. Their $1-\sigma$ error bars are derived from the fluctuations between the MD-Patchy mocks. The dashed-dotted line indicates the ratio of the linear growth factors of our $\Lambda$CDM model.}
\label{fig:growcomp}
\end{figure}

\cref{fig:etah_data} clearly suggests that the MF values are larger at lower redshift.  \cref{fig:ehper} shows the percentage growth from the higher redshift CMASS bin to the lower redshift LOWZ for a selection of 10 of our 24 density values.  This shows that the growth is fairly constant as a function of density, with a mild decrease as the density increases.  Importantly, the percentage growth of 20--25\% in the MF is about a factor of two larger than the $\approx 10\%$ growth in the two-point correlation function between these two samples.

The percentage growth values for the data in the NGC and SGC seem to agree with the mocks relatively well. Barring the limit of low density, and the case of $\eta_{0}^{h}$, the values are close to, if not within, the $1-\sigma$ error bars. However, it is clear that the growth for the data is systematically higher than the mean growth of the mocks. 

\subsection{Model comparison}\label{subsec:model}

Having detected significant growth in the higher-order contributions to the MFs, we trace its origin in this section. To this end we analyzed a set of 16 N-body simulations of the same Planck cosmology, but with varying phases \citep{2016MNRAS.461.4125G}.
For these simulations we determined the two-point correlation functions and MFs for the underlying dark matter field and for the galaxies derived from a self-adjusted HOD model at two different redshifts. The redshifts that we use correspond to a snapshot of the simulation at $z=0.3$ for our LOWZ bin and $z=0.5$ for our CMASS bin.

The HOD model we fit is of the form used in \cite{2009ApJ...707..554Z} and \cite{2015ApJ...810...35K}. We derive its five parameters using code developed by \cite{2017MNRAS.472..577Y}, performing a fit to our measured redshift space two-point correlation function monopole of the data. More specifically, we use the FoF halos provided by the simulation, select them with the HOD prescription, and redshift the final selection in the z-direction. We used this approach instead of the standard one, as parameters derived from a fit to the projected two-point function of the data as found in the literature (although based on a different reference cosmological model) did not give a good match for the amplitude of the redshift space two-point monopole. For LOWZ for example, we needed a slightly higher $M_{\rm{cut}}$ and lower $\sigma$ than the best fit of \cite{2013MNRAS.429...98P}. However, the final parameters of our model lie well within their ``Mean Full'' error bars.

The philosophy of using the redshift space two-point correlation function of the data that we also used in the rest of our analysis was to generate a galaxy distribution from the HOD model that best reflects the two point properties we measure. Every deviation in the higher-order only MFs of the resulting distribution from those of the data is then maximally informative of the properties of the higher orders.

\begin{figure*}
\begin{center}
\includegraphics[width=0.9\textwidth]{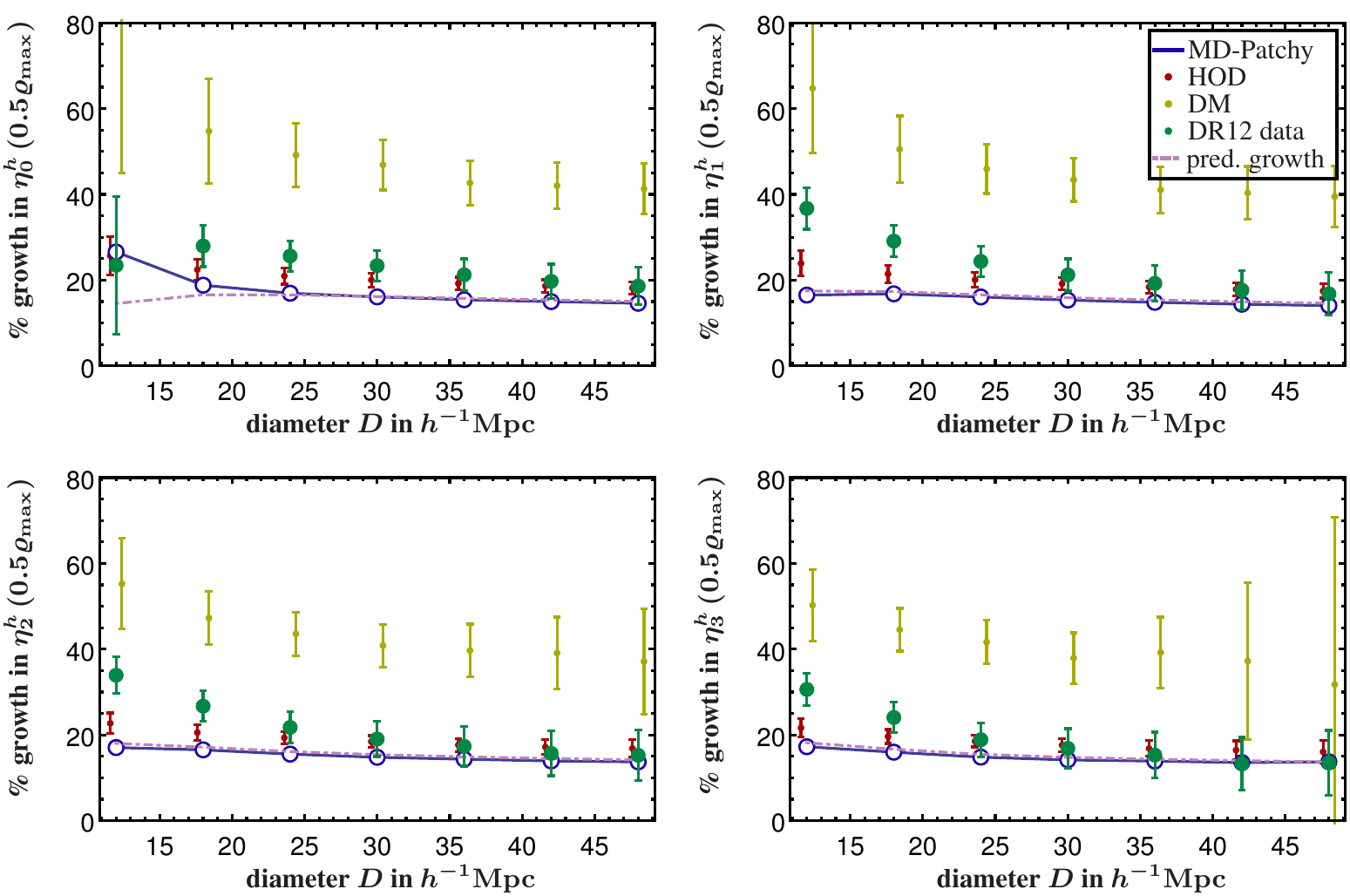}
\end{center}
\caption{Comparison of the growth in the higher-order correlations, measured through $\eta^{h}_{\mu}$, between our two redshift bins for different cases. We use only data from the NGC here. Blue hollow points joined by the line show the mean of the MD-Patchy mocks. The red points are derived from the galaxy samples of our own HOD model, and the yellow points show the growth of the underlying dark matter field. In those two cases the $1-\sigma$ error bars are derived from realizations with different phases of the initial modes. The green solid points are the values measured for the galaxies from the final DR 12 sample. Their $1-\sigma$ error bars are derived from the fluctuations between the MD-Patchy mocks. The dashed-dotted line shows the expected growth when scaling all higher-order functions of the MD-Patchy mocks by their corresponding powers of the linear amplitude growth (see text). $\eta^{h}_{\mu}$ is measured at $50\%$ of the maximal density, i.e. at $1.07\times10^{-4}\hDens$, but density dependence is relatively modest for large scales, cf. \cref{fig:ehper}.
}
\label{fig:growcomphigh}
\end{figure*}

\cref{fig:growcomp} shows the growth of the two-point correlation function between our CMASS and our LOWZ bin. The growth is largest for the underlying dark matter field in yellow which agrees well with the linear growth of the amplitude of the dark matter density field given by the ratio of their linear growth factors $D(z=0.3)/D(z=0.5) \approx 1.228$.

From the comparison of the final HOD galaxy two-point correlations to those of their dark matter field, we find that the approximate linear bias drops from $b_{1}^{2}=4.93$ to $b_{1}^{2}=4.60$. This $7\%$ decrease in the correlation amplitude counteracts the increase of the clustering from the non-linear evolution of the dark matter density field and lets the red points in \cref{fig:growcomp} lie below those for the underlying dark matter field.

The growth derived from the data shown as the green filled circles lies in the expected range for both HOD galaxies (in red) and mock galaxies from the MD-Patchy mocks (in blue), except at very small scales. For MD-Patchy this may be related to the fundamental resolution of the underlying density field from which the mocks were created. For our HOD model it probably is related to the use of FoF halos and the resolution of light dark matter halos.

\cref{fig:growcomphigh} shows the growth of the higher-order contribution to the MFs. It follows the general behavior of the two point case in \cref{fig:growcomp}, with the strongest growth being realized in the dark matter field. Also here the data lies within the range of the simulated galaxies, except at the smallest scales. Again, the scale dependence of the growth is fairly weak, but more important than in the two-point case.

\begin{table}
\centering
\setlength\tabcolsep{3pt}
\caption{Significance of growth with redshift in each galactic cap. The $\chi^{2}$ values of significance are for the growth in the MFs for the data between the CMASS and LOWZ redshift bins for $\eta_{\mu}^h$ in each galactic cap at $\vr = 0.5\varrho_{\mathrm{max}}$ for 7 degrees of freedom. These values quantify the growth shown in \cref{fig:growcomphigh}.}
\label{tab:chisq_growth}
\begin{tabular}{lcccc}
\hline \hline 
$\mu$ & \multicolumn{2}{c}{NGC} & \multicolumn{2}{c}{SGC}\\
\hline
 & $\chi^{2}$ &  $p$-value & $\chi^{2}$ &  $p$-value \\
\hline
0 & 96.4 & $10^{-18}$ & 13.4 & 0.06 \\
1 & 111.0 & $10^{-20}$ & 16.1 & 0.02 \\
2 & 108.4 & $10^{-20}$ & 15.2 & 0.03 \\
3 & 106.0 & $10^{-19}$ & 14.0 & 0.05 \\
\hline
\end{tabular}
\end{table}

For the data shown in \cref{fig:growcomphigh} we quantify the significance with which we measure the growth in the higher orders in \cref{tab:chisq_growth}. We determine the $\chi^2$ values by computing the difference of the values of $\eta^{h}_{\mu}$ in our CMASS and LOWZ bins for the seven radii plotted in \cref{fig:growcomphigh}. As we determine the significance for the difference of two independent values, we use the inverse of the sum of the covariance matrices of the CMASS and LOWZ data for obtaining the $\chi^{2}$. The data used for $\eta^{h}_{3}$ can be found in \cref{tab:data} at the end of the text. We do not use the fraction shown in \cref{fig:growcomphigh} directly, as for a ratio distribution the calculation  of the significance is less straightforward than for the difference. We find a very significant detection of growth. Even if we remove the first two radial points for which it is not entirely clear how well the mocks and their fluctuations describe the data, we are still left with a $5.5\sigma$ detection. Due to the strong correlation in $\vr$, adding in more points in $\vr$ does not change this picture. We investigated a compression of the $\vr$ direction into a single value by a transformation into the eigenbasis of the covariance matrix. The leading two eigenvectors give a good description of the data as a function of density and using them instead of only the value at $\vr = 0.5\varrho_{\mathrm{max}}$ as in \cref{tab:chisq_growth} does not increase the significance by much. 

In order to test if we understand the growth of the higher orders, we added a line to \cref{fig:growcomphigh} that shows the predicted growth for the mean of the mocks derived from the growth of their two-point function. This has been calculated following the observation of the \hyperlink{txt:dr12paper}{DR12 paper} that in a perturbative approach to the higher-order connected correlation functions the linear power spectrum enters the series expansion in \cref{eq:PowerSeriesDecomp} with the same power as the density. So the leading effect of increasing the linear power spectrum amplitude can be obtained by a simple rescaling of the density. We derive this rescaling factor at each radius from the measured integrated two-point correlation function. Applying this correction to the higher-order terms as well, by evaluating the mean MD-Patchy CMASS $\eta^{h}_{\mu}$ at the shifted $\vr$ position and taking the ratio, gives the dashed-dotted line in \cref{fig:growcomphigh}. The good match with the measured growth in the mocks indicates that it is indeed the leading terms at each correlation order that grow proportionally to their respective amplitude factor. This is true within the errors also for the data (their predicted growth is not shown on the plot for clarity) and illustrates how structure grows hierarchically.

Finally, we note that the reasonable agreement in the shape of the growth for our HOD galaxies and the MD-Patchy mocks provides a basic test of the influence of the boundaries. As the HOD galaxies are derived from a boundaryless periodic box and the MD-Patchy mocks include all effects of the our boundary treatment, we can be confident that the main conclusions of this paper regarding the detection of growth in the higher orders are not affected by possibly remaining issues with the mask.

\section{Summary \& Conclusions}\label{sec:SC}

In this paper, we perform an unprecedented analysis of higher-order correlation information on the largest spectroscopic redshift survey to date. We use Minkowski Functional analysis along with independently calculated two-point correlation functions to measure non-Gaussianity in large-scale structure. Our detection of non-Gaussianity is highly significant, with $\chi^{2}$ values of $\mathcal{O}(10^{3})$ for $24$ degrees of freedom across the NGC and SGC of the SDSS-III DR12.

We measure the percentage growth in redshift of the higher-order part of the MFs for the first time. This growth is measured to be $15\%-35\%$, depending on functional, density and scale. We show the redshift evolution of non-Gaussianity is greater in the higher-order contribution to the MFs than in the two-point term. When comparing the integrated two-point correlations to the higher-order part of the MFs, the excess of the higher orders approaches a factor of $2$ on small scales.

Using simulations we show that the growth of higher-order structure of the underlying dark matter field is even larger, exceeding $60\%$  on small scales for some of the functionals, but that it is counterbalanced in the galaxy field by a decrease in the bias which reduces the growth of the two-point amplitude by $7\%$.

Finally we show how measuring the higher-order MFs as a function of density allows us to perform a simple test of hierarchical clustering. Rescaling the density corresponds to increasing the correlation strength in the higher-orders proportional to their order and reproduces the results for the growth very well.

Our results motivate multiple avenues for future analysis of higher-order correlations, including measurement of the growth of the integrated three-point and four-point functions. The growth in the higher-order correlations may also be applied in relation to the linear growth rate.

\section*{Acknowledgements}
AW acknowledges support from the German research organization
DFG, Grants No.~WI 4501/1--1 and No.~WI 4501/2--1.
DJE acknowledges support from National Science Foundation grant AST-1313285, from U.S. Department of Energy grant DE-SC0013718, and as a Simons Foundation Investigator.

The SAO REU program is funded by the National Science Foundation REU and Department of Defense ASSURE programs under NSF Grant AST-1659473, and by the Smithsonian Institution. This research has also made use of results from NASA's Astrophysics Data System. 

\begin{table*}
\centering
\setlength\tabcolsep{3pt}
\begin{tabular}{c|ccccccc}
\hline \hline
 $D$ & 12 & 18 & 24 & 30 & 36 & 42 & 48 \\ \hline
 $\eta^{h}_{3,\mathrm{d}}$ & 0.262 & 0.602 & 0.967 & 1.338 & 1.689 & 2.005 & 2.287 \\ \hline \hline
 $\eta^{h}_{3,\mathrm{m}}$ & 0.236 & 0.570 & 0.954 & 1.343 & 1.710 & 2.045 & 2.351 \\ \hline
 $\sigma$ & 0.0055 & 0.013 & 0.026 & 0.043 & 0.065 & 0.091 & 0.124 \\ \hline
 12 & 1 & 0.9 & 0.797 & 0.711 & 0.653 & 0.599 & 0.545 \\
 18 & $\ldots$ & 1 & 0.953 & 0.88 & 0.815 & 0.742 & 0.663 \\
 24 & $\ldots$ & $\ldots$ & 1 & 0.967 & 0.91 & 0.834 & 0.74 \\
 30 & $\ldots$ & $\ldots$ & $\ldots$ & 1 & 0.972 & 0.909 & 0.811 \\
 36 & $\ldots$ & $\ldots$ & $\ldots$ & $\ldots$ & 1 & 0.969 & 0.884 \\
 42 & $\ldots$ & $\ldots$ & $\ldots$ & $\ldots$ & $\ldots$ & 1 & 0.95 \\
 48 & $\ldots$ & $\ldots$ & $\ldots$ & $\ldots$ & $\ldots$ & $\ldots$ & 1 \\
 \hline
\end{tabular}
\begin{tabular}{c|ccccccc}
\hline \hline 
 $D$ & 12 & 18 & 24 & 30 & 36 & 42 & 48 \\ \hline
 $\eta^{h}_{3,\mathrm{d}}$ & 0.199 & 0.482 & 0.806 & 1.135 & 1.452 & 1.755 & 1.997 \\ \hline \hline
 $\eta^{h}_{3,\mathrm{m}}$ & 0.200& 0.487 & 0.823 & 1.165 & 1.488 & 1.785 & 2.051 \\ \hline
 $\sigma$ & 0.0038 & 0.0093 & 0.018 & 0.030 & 0.045 & 0.065 & 0.09 \\ \hline
 12 & 1 & 0.89 & 0.743 & 0.636 & 0.556 & 0.489 & 0.41 \\
 18 & $\ldots$ & 1 & 0.937 & 0.851 & 0.769 & 0.696 & 0.598 \\
 24 & $\ldots$ & $\ldots$ & 1 & 0.965 & 0.902 & 0.828 & 0.721 \\
 30 & $\ldots$ & $\ldots$ & $\ldots$ & 1 & 0.973 & 0.914 & 0.813 \\
 36 & $\ldots$ & $\ldots$ & $\ldots$ & $\ldots$ & 1 & 0.969 & 0.882 \\
 42 & $\ldots$ & $\ldots$ & $\ldots$ & $\ldots$ & $\ldots$ & 1 & 0.948 \\
 48 & $\ldots$ & $\ldots$ & $\ldots$ & $\ldots$ & $\ldots$ & $\ldots$ & 1 \\
 \hline
\end{tabular}
\caption{Extracted higher-order part of the MF for the data $\eta^{h}_{3,\mathrm{d}}$ and the MD-Patchy mocks $\eta^{h}_{3,\mathrm{m}}$ using NGC only as a function of diameter $D$ at $\vr=0.5\varrho_{\mathrm{max}}$, i.e. $1.07\times10^{-4}\hDens$. Left table for our LOWZ bin, right table for our CMASS bin. The large tables show the symmetric correlation matrices, with the values on the left indicating the respective diameter of the row. $\sigma^2$ gives the diagonal elements of the covariance matrix and $\sigma\otimes\sigma$ multiplies the correlation matrix to give the covariance matrix. The covariances have been derived from 399 MD-Patchy mocks for each of the tables. Dividing the $\eta^{h}_{3,\mathrm{d}}$ values on the left by those on the right gives the green data line in \cref{fig:growcomp} and the covariances can be used to derive those of the MD-Patchy mocks shown in the same figure. As the different density values are strongly correlated and the same is true for the different functionals, the one set of tables presented here already contains a large fraction of the information contained in our measurements.}
\label{tab:data}
\end{table*}

Funding for SDSS-III has been provided by the Alfred P. Sloan
Foundation, the Participating Institutions, the National Science
Foundation and the US Department of Energy Office of Science.
The SDSS-III website is http://www.sdss3.org/.

SDSS-III is managed by the Astrophysical Research Consortium
for the Participating Institutions of the SDSS-III Collaboration
including the University of Arizona, the Brazilian Participation
Group, Brookhaven National Laboratory, Carnegie Mellon University,
University of Florida, the French Participation Group, the
German Participation Group, Harvard University, the Instituto de
Astrofisica de Canarias, the Michigan State/Notre Dame/JINA Participation
Group, Johns Hopkins University, Lawrence Berkeley
National Laboratory, Max Planck Institute for Astrophysics, Max
Planck Institute for Extraterrestrial Physics, New Mexico State University,
New York University, Ohio State University, Pennsylvania State University, University of Portsmouth, Princeton University,
the Spanish Participation Group, University of Tokyo, University
of Utah, Vanderbilt University, University of Virginia, University
of Washington and Yale University.

\bibliographystyle{mnras}
\bibliography{references}

\bsp	
\label{lastpage}
\end{document}